\documentclass[final,1p,times,authoryear]{elsarticle}
\usepackage{longtable}
\usepackage{graphicx}
\usepackage{amsmath,amsfonts,amssymb,amstext}
\usepackage{wrapfig}
\usepackage{multirow}
\usepackage{epsfig}
\usepackage{psfrag}
\usepackage{subfigure}
\usepackage{xcolor}
\usepackage{mathrsfs}
\usepackage{stackengine}
\usepackage{float}
\usepackage{tabularx}
\usepackage{natbib}
\usepackage{epstopdf}
\usepackage{caption} 
\usepackage{subcaption}

\usepackage{tablefootnote}

\usepackage[flushleft]{threeparttable}

\journal{New Astronomy}

\begin{document}

\begin{frontmatter}

\title{X-ray reprocessing: Through the eclipse spectra of low mass X-ray binaries with \textit{XMM-Newton}}

\author[a]{Nafisa Aftab\corref{cor1}}
\cortext[cor1]{Corresponding Author}
\ead{aftabnafisa@gmail.com}
\author[a]{Biswajit Paul}

\address[a]{Raman Research Institute, C V Raman Avenue, Sadashivanagar, Bangalore 560080, India}

 \begin{abstract}
Eclipsing X-ray binaries make an ideal
condition to study reprocessed X-rays, as the X-rays detected during eclipse are purely reprocessed while
the much brighter primary X-rays are blocked by the companion star.
We carried out a comprehensive study of X-ray reprocessing with four eclipsing Low Mass X-ray Binary (LMXB)
systems by comparing X-ray spectra during and outside eclipse with \textit{XMM-Newton} EPIC pn observations. 
The 17 observations of MXB 1659$-$298, AX J1745.6$-$2901, EXO 0748$-$676 and XTE J1710$-$281  give unique features of the systems. For example, X-ray reprocessing characteristics in AX J1745.6$-$2901  is found to be nearly same irrespective of the intensity state; there is an indication of different types of variable warp structures in the inner accretion disk in EXO 0748$-$676,
a high out-of-eclipse to eclipse flux ratio in XTE J1710$-$281 inspite of a large size of the accretion disk 
perhaps indicates low scale height of the accretion disk. The eclipse spectra for some of the LMXB sources are  reported for the first time. We have derived the fractional visible area of the accretion disk during maximum eclipse phase for various obscuration geometries. The out-of-eclipse to eclipse flux ratio in LMXBs observed to be smaller compared to that found in High Mass X-ray Binaries. This indicates greater reprocessing in LMXBs despite having less dense, less extended stellar wind from the companion. The X-ray reprocessing efficiencies observed in LMXBs indicate large dependencies of X-ray reprocessing on the scale height of the accretion disk, relative size of the disk compared to the companion and some other unknown factors.
 \end{abstract}


 \end{frontmatter}
% % ===========================================
% % =====================================================
% % ==================================================================
% % % % =========================================================================================================================================================================
\section{Introduction}
In X-ray binaries, most of the X-rays originate in the inner accretion disk
(in case of disk accretion) and boundary layer of the neutron star and
from the pole of the neutron star in case of pulsars.
These X-rays are called primary X-rays. Some of these X-rays come out
directly. A significant fraction of the primary X-rays interacts with the surrounding matter 
and are re-emitted in different directions. This phenomenon is known as X-ray reprocessing and 
the re-emitted X-rays are called reprocessed X-rays. Depending upon the type of interaction, 
the reprocessed or secondary emission could be X-ray, UV and / or optical \citep{Aftab2019}.
% % % -----------------------------------------------------------------------
Reprocessing of X-rays have been observed in active galactic nuclei (AGN), accreting systems
with black hole, neutron star, white dwarf as compact objects (\citealt{1994Zycki}, \citealt{2009Gierlinski}, \citealt{1996Jong}, \citealt{2003Suleimanov}).
\par
In this work we focus on X-ray reprocessing characteristics of Low Mass X-ray Binary (LMXB) systems. 
 In LMXBs the companion star has a mass 
 M $\leq$ M$_{\odot}$ as donor star and either black hole or neutron star as compact object. The 
 companion is often a late type star (spectral type A or later i.e K, L, M types) or a white dwarf in a few cases. 
In LMXB systems the main reprocessing agent is the accretion disk and any structure on the disk if present.
The observability X-ray reprocessing characteristics in LMXBs depend upon several factors, for example
i) the scale height of the accretion disk ii)  inclination of the system with respect to our line of sight
iii) the size of the accretion disk with respect to the size of the companion iv) structures on the accretion disk 
v) distribution of matter in the system and their densities vi) the chemical composition and ionization levels
of the matter around the compact object vii) wind outflow from the accretion disk or accretion disk atmosphere
viii) orbital phase of the system. The scale hight of the accretion disk determines the solid angle covered 
by the accretion disk surface to central source, hence how much central radiation will be reprocessed.
The inclination determines how much fraction of the central radiation will be scattered towards us.
\par
Study of X-ray reprocessing gives us an understanding of the environment that causes the 
reprocessing of X-rays. But the main challenge in this study is that,  
the detected X-rays are a mixture of primary and reprocessed X-rays and the primary 
 X-rays are much brighter than the reprocessed X-rays. However during  eclipse, the primary X-rays
 from the compact object are completely
 blocked by the companion and the received X-rays are the reprocessed
 X-rays only, which is free of primary emission. Hence X-rays received during eclipses from the eclipsing LMXBs
 are ideal to study the X-ray reprocessing in these systems. The portion of the accretion disk visible to us during eclipse depends upon the relative size of the disk compared to the size of the companion, inclination of the system with respect to the line of sight and the orbital separation.
%  
% --------------------------
% 
\par
In this work we have analyzed four LMXB systems (MXB 1659$-$298, EXO 0748$-$676, AX J1745.6$-$2901 and XTE 1710$-$281)
during eclipse and out-of-eclipse phases with {\it XMM-Newton} EPIC pn.
The transient, bursting and dipping neutron star LMXB system MXB 1659$-$298 was 
discovered with \textit{SAS-3} \citep{1976Lewin}. 
The source is an eclipsing system \citep{1984Cominsky} with an orbital period of 7.116 hr \citep{1989Cominsky}.
\citet{2017Jain_1659} suggests a massive third body around the binary orbit with mid eclipse time analysis of the source
with \textit{RXTE} and \textit{XMM-Newton}.
EXO 0748$-$676, a neutron star LMXB system was discovered with \textit{EXOSAT} in 1985 and was found to have an orbital period
of 3.824 hour and eclipses lasting for 
8.3 min \citep{1985Parmar}. After 24 years of continuous accretion, the source went into quiescence
in 2008 \citep{2011Trigo_0748}. \citet{2017Gayathri} have inferred a strong evidence of accretion disk evolution
at time scale of few hours, from the modulation of optical lightcurve of the same \textit{XMM-Newton} observations we 
have analyzed in this work. AX J1745.6$-$2901 is an eclipsing neutron star LMXB system 1.45 arcmin from Sgr A$^{\star}$,
was discovered by ASCA during 1993-1994 outburst with an orbital period of 
$\sim$8.356 hours \citep{1996Maeda}. \textit{XMM-Newton} and \textit{Chandra} data showed evidence of
different dust layers surrounding the source \citep{2017Jin, 2018Jin}, which causes asymmetric shape of the lightcurves 
during eclipses. 
XTE J1710$-$281 is a transient LMXB system \citep{1998Markwardt} serendipitously detected with \textit{RXTE}-PCA.
The source shows eclipse for an average duration of $\sim$420 s  in each orbit with an orbital period of 
3.281 hr \citep{2011Jain}. \citet{2018raman} have found a signature of accretion disk winds in this 
source through the observation of highly ionized
Fe lines with \textit{Suzaku}, they have  also observed evolution of dips in each orbit with \textit{Chandra} data.
\par
Eclipse spectrum of some of the sources have been studied before (MXB 1659$-$298: \citealt{2001Sidoli_MXB}, 
EXO 0748$-$676: \citealt{2018Psaradaki}) but here we report results from first comprehensive study
of LMXB eclipse and out-of-eclipse spectra with {\it XMM-Newton} EPIC pn. 
A similar investigation of the X-ray reprocessing in High Mass X-ray Binaries (HMXBs) has been reported in \citet{Aftab2019}.
Here we compare the eclipse flux with the out-of-eclipse flux and see the reprocessing efficiency of LMXBs as
a class and make a comparison with HMXBs. Very detailed analysis of individual sources are not discussed here.
We also aim to study the variability in X-ray reprocessing and its possible dependencies on the system characteristics
to investigate the stability of reprocessing in a given source with multiple data sets. 
A list of known orbital parameters of these sources which are relevant to this work are given in Table \ref{intro}.
We have estimated the binary separation (a) of each source using Kepler's third law of motion,
4$\pi^{2}\rm a^{3}$ = GMP$^{2}$  \citep{Frank_etal_02}. Where M is the total mass of the neutron star and the companion star,  P is the binary period, G is the gravitational constant. We have further estimated the outer radius of the accretion disk (R$_{\rm out}$) using the relation 
R$_{\rm out}$ $\sim$0.8$\times$R$_{\rm LNS}$, where R$_{\rm LNS}$ is the Roche lobe radius of the neutron star
 R$_{\rm LNS}$ = a $\times$ $\frac{0.49 \rm q^{-2/3}}{0.6 \rm q^{-2/3} + ln(1 +  \rm q^{-1/3})}$  \citep{Frank_etal_02}, 
 q is the ratio of the companion star's mass with neutron star's mass.
\par
The manuscript is organized as follows. In $\S$2 we give details of the observations and data reduction procedure
with {\it XMM-Newton} EPIC pn, in this section we describe the spectral anlalysis of the eclipse and out-of eclipse
events. We discuss the results from the comparison of the eclipse
and out-of-eclipse spectra of each observation and also from the comparison of the eclipse (and out-of-eclipse spectra) spectra  at 
different epoch of the same source in $\S$3, here we compare the X-ray reprocessing between LMXBs and HMXBs.
 Finally we conclude the results in $\S$4.
% % % 
% ---------------------------------
\begin{table*}[ht]
%  \tiny
\small
% \begin{center}
\label{table}
% \vspace{1ex}
\caption{List of basic parameters of the eclipsing LMXBs analyzed in this work, where $\mathrm{P_{orb}}$: 
Orbital period (Hr), q: Mass ratio (Ratio of companion star's mass with the mass of neutron star) M$_{\rm C}$: Mass of the companion (M$_{\odot}$), $i$: inclination of the orbit with respect to the line of sight  ($^\circ$),
d: Distance from Earth (kpc) (M$_{\odot}$, R$_{\odot}$: Mass and radius of Sun respectively)}
\centering
% \begin{small}
\begin{tabular}{l lr rrr }
\hline
% ---------------------------------
% ---------------------------------
  Source	       & $\mathrm{P_{orb}}$ & q & M$_{\rm C}$    & i   &   d   \\
  
        &    (Hr)   &  & (M$_{\odot}$)      &   ($^\circ$)     &    (kpc)    \\
  	               &                                                  &   &                      &            &    \\

		\hline
MXB 1659$-$298	   & 7.1161099(3)$^{a_{1}}$$^{,}$ $^{a_{2}}$  &  0.35-0.95\textcolor{magenta}{$^\ddagger$}       & (0.9$\pm$0.3)$^{a_{2}}$        & (72$\pm$3)$^{a_{2}}$ & (10-16)$^{a_{3}}$  \\
   & & &       & \\
		\hline
EXO 0748$-$676	   & 3.82410725(75)$^{b_{1}}$  & 0.222$^{+0.07}_{-0.08}$$^{b_{2}}$           & 0.44$^{b_{2}}$                & 76.5$^{+1.4}_{-1.1}$$^{b_{2}}$            & (5-11)$^{a_{3}}$    \\
   & &  &       &  \\
		\hline
AX J1745.6$-$2901  & 8.3510389$\pm$0.0002055   $^{c_{1i}}$$^{,}$ $^{c_{1j}}$   &  $\leq$ 0.6$^{c_{2}}$         & $\leq$ 0.8$^{c_{2}}$            & 70$^{+7}_{-15}$$^{c_{3}}$    & 8.3$^{c_{4}}$   \\
   & & & &       &   \\
		\hline
XTE J1710$-$281	   & 3.281063218(5)$^{d_{1}}$$^{,}$ $^{d_{4}}$     &  0.3$^{d_{5}}$     & 0.42\textcolor{magenta}{$^{\clubsuit}$}            & (75-80)$^{d_{2}}$            & (12-16)$^{d_{3}}$  \\
   & & & &       &  \\
		\hline
% --------------------------- 
 \end{tabular}
%  \end{small}
\label{intro} \\
% \end{center}
% \\
\begin{tablenotes}
\item[*] 
$^{a_{1}}$\citep{1989Cominsky}, $^{a_{2}}$\citep{2018Iaria}, $^{a_{3}}$\citep{2006Trigo}\\
$^{b_{1}}$\citep{1986Parmar_etalPpaper1},
% % % $^{b2}$\citep{2009_Munozetal},
$^{b_{2}}$\citep{Knight_etal2022}\\
$^{c_{1i}}$\citep{1996Maeda}, $^{c_{1j}}$\citep{2008_Hyodo}, $^{c_{2}}$\citep{2017Ponti}, $^{c_{3}}$\citep{2018Ponti}, $^{c_{4}}$(\citealt{Genzel_etal_10}, \citealt{Bland_Gerhard_16}, \citealt{Gillessen_etal_2017}) \\
$^{d_{1}}$\citep{2011Jain}, $^{d_{2}}$\citep{Frank_etal_87}, $^{d_{3}}$\citep{Galloway2008}, $^{d_{4}}$\citep{2022Jain_1710}, $^{d_{5}}$\citep{2017Gayathri} \\
% % % $^{\star}$Derived from orbital period ($\frac{\rm R_{\rm C}}{\rm R_{\odot}}$ $\sim$0.11P$_{\rm orb}$, \citealt{Frank_etal_02}),
\textcolor{magenta}{$^{\clubsuit}$}Derived from $\frac{\rm M_{\rm C}}{\rm M_{\rm NS}}$ $\sim$0.3,  M$_{\rm NS}$ $\sim$1.4M$_{\odot}$ \citep{2017Gayathri}\\
\textcolor{magenta}{$^\ddagger$}Derived from M$_{\rm C}$ = (0.9 $\pm$ 0.3)M$_{\odot}$ \citep{2018Iaria} and M$_{\rm NS}$ = (1.48 $\pm$ 0.2)M$_{\odot}$\citep{Ozel_etal2012}
\end{tablenotes}
\end{table*} 
% % % 
\section{Observation and data analysis}
\textit{XMM-Newton} observatory consists of 3 sets of co-aligned X-ray telescopes each with an effective area of 1500 cm$^{2}$,
launched in December 1999.
 At the focal plane of each telescope three European Photon Imaging Cameras (EPIC) are placed, two of which are made of Metal 
 Oxide Semi-conductor (MOS) CCDs \citep{2001turn} and one  uses pn CCDs  \citep{2001struder}.
% % %  
% % %  
 EPIC pn instrument operates  in the energy range
 of 0.15--15 keV and provides spectral 
 resolution of 20--50 (E/$\Delta$E) and angular resolution of 6$''$ over the 30$'$ 
 field of view. EPIC pn provides different read out modes,  
 namely FullFrame, LargeWindow, SmallWindow, Timing mode \citep{2012lumb}. 
 According to the science requirements any of these modes can be selected.
\par
We referred to  \citet{2007Liu}  for eclipsing LMXBs and then searched for the available EPIC pn observations
in HEASARC. We found eclipse observations of four LMXBs. From these we selected a total of 17 observations, 
 which were carried out in the Imaging mode. 
 The log of observations along with 
 the mode of operations are given in Table \ref{mode}.
 We  used the \textit{XMM-Newton} Science Analysis Software (SAS) version
14.0.0 to reduce the data. The event files were extracted with the
tool {\small  EVSELECT}. Particle background were checked
and no flares have been found in any of the observation. Events were extracted 
in the energy range of 0.3-12 keV. Pileup were checked with the  
task {\small EPATPLOT}
and no pile up effect were noticed for the chosen energy range (0.3-10 keV).
% **************************************************************************************************
\par
\begin{figure*}
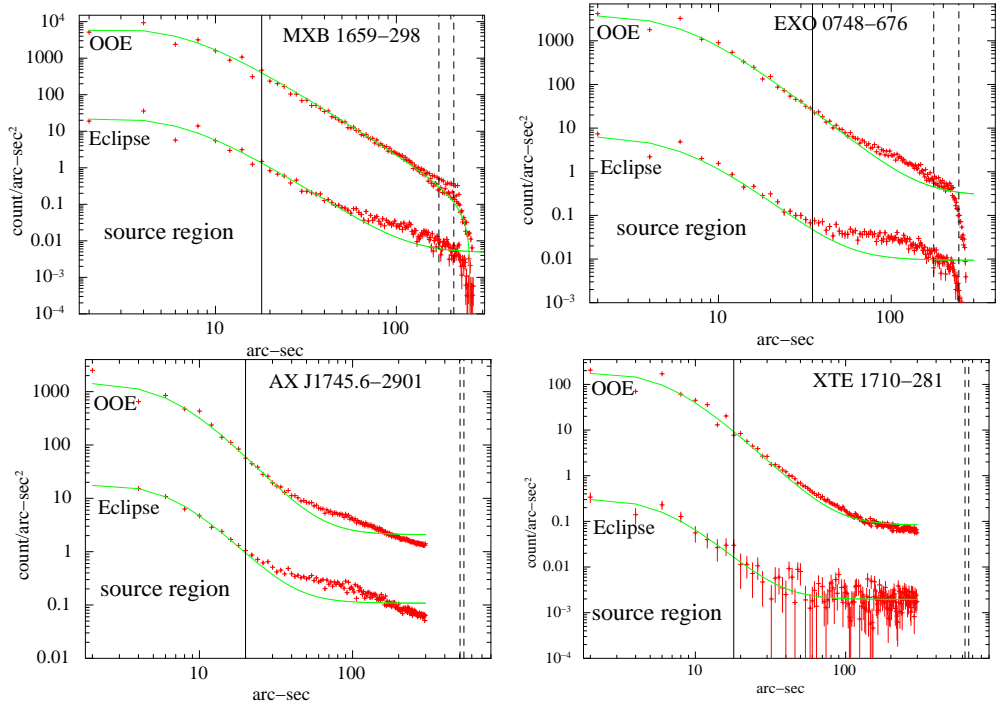

 \centering
 \includegraphics[scale=0.25, angle=-90]{combine_rad_prof_MXB_701_3.ps}
 \includegraphics[scale=0.25, angle=-90]{combine_rad_prof_EXO_801_3.ps}
 \includegraphics[scale=0.25, angle=-90]{combine_rad_prof_AXJ_4.ps}
 \includegraphics[scale=0.25, angle=-90]{combine_rad_prof_XTE_5.ps}
 \caption{The radial profiles of  MXB 1659$-$298 (top left), EXO 0748$-$676 (top right), AX J1745.6$-$2901 
 (bottom left) and XTE J1710$-$281 (bottom right)  during eclipse (lower plots) and out-of-eclipse 
 (upper plots) phase respectively. One radial profile is shown for each source.
 The X and Y axis shows the distance from the source in arc-sec and counts/arc-sec$^{2}$ received in the
 whole observation respectively. Zero value in the X axis indicates the position of 
 the source centroid. The red points denote radial profile of the source in the field of view, where the
 green lines denote fit of a point spread function to the data.
 The region between Y axis and the solid line shows the source region and that between the two dashed lines
 describes background region.}
 \label{radprof}
 \end{figure*}

We extracted images during eclipse and out-of-eclipse (OOE) of all the observations of the sources analyzed in this work.
With the 
tool {\small  RADPROF} and using the image and event files we obtained radial profiles of the sources of
 all the observations during eclipse and OOE phase. Figure \ref{radprof} shows examples of one radial profile
of each of the source. The radial profiles of  MXB 1659$-$298 (OB ID: 0008620701), EXO 0748$-$676 (OB ID: 0160760801), 
 AX J1745.6$-$2901 (OB ID: 0402430301) and XTE J1710$-$281 (OB ID: 0206990401) are shown in this figure
 in top left, top right, bottom left and bottom right panel respectively.
 The X axis shows the distance from the source in arc-sec, where zero value in this axis indicates the position of the source centroid. 
 The Y axis denotes counts receivded per arc-sec$^{2}$ during the whole observation.
 The red points and green lines denote radial profile of the source in the field of view and fit of a point spread function
 (from the calibration files) to it respectively. The region between the Y axis and the solid line shows the source
 region and that between the two dashed lines shows background region used for the extraction of the source and background spectra 
 for spectral analysis.
This profile helps us to test the presence of excess emission which may be scattered from interstellar dust
(during the phase for which the profile has been generated) into the line of sight. Examining the plots we observe that excess emissions are present
in a few observations, but are not very close to the source. These excess emissions are present at a certain range of distances 
from the sources. We extracted circular source region
 from   regions surrounding the source centroid and then extracted the background region of same size of the source, 
 from a region which are free from any excess emission along the line of sight and also are free from any X-ray sources.  
 Seeing the image quality we have chosen the size of the source 
 regions.
For example we have extracted the source regions with a radius of
18$''$, 35$''$, 20$''$ and 18$''$ for MXB 1659$-$298, EXO 0748$-$676, AX J1745.6$-$2901 and XTE J1710$-$281 respectively, where any excess emissions have not been observed.
Similarly we have chosen the background regions from
regions which are also free from excess emissions and are far away from the source. 
Edge of a CCD chip has been avoided for the extraction of the regions. Diffused 
emissions get eliminated with the background subtraction while obtaining the lightcurves and spectra.
%  ***************************************************************************************************
% 
 We extracted the source and background lightcurves from the single and double events (with PATTERN$\leq$4)  
 for the full exposure time with the 
 task {\small  EVSELECT}. 
 We then obtained background corrected source lightcurve with the 
 task {\small  LCCORR}.
 \subsection{EVENT SELECTION FROM THE LIGHTCURVES}
\begin{figure*}
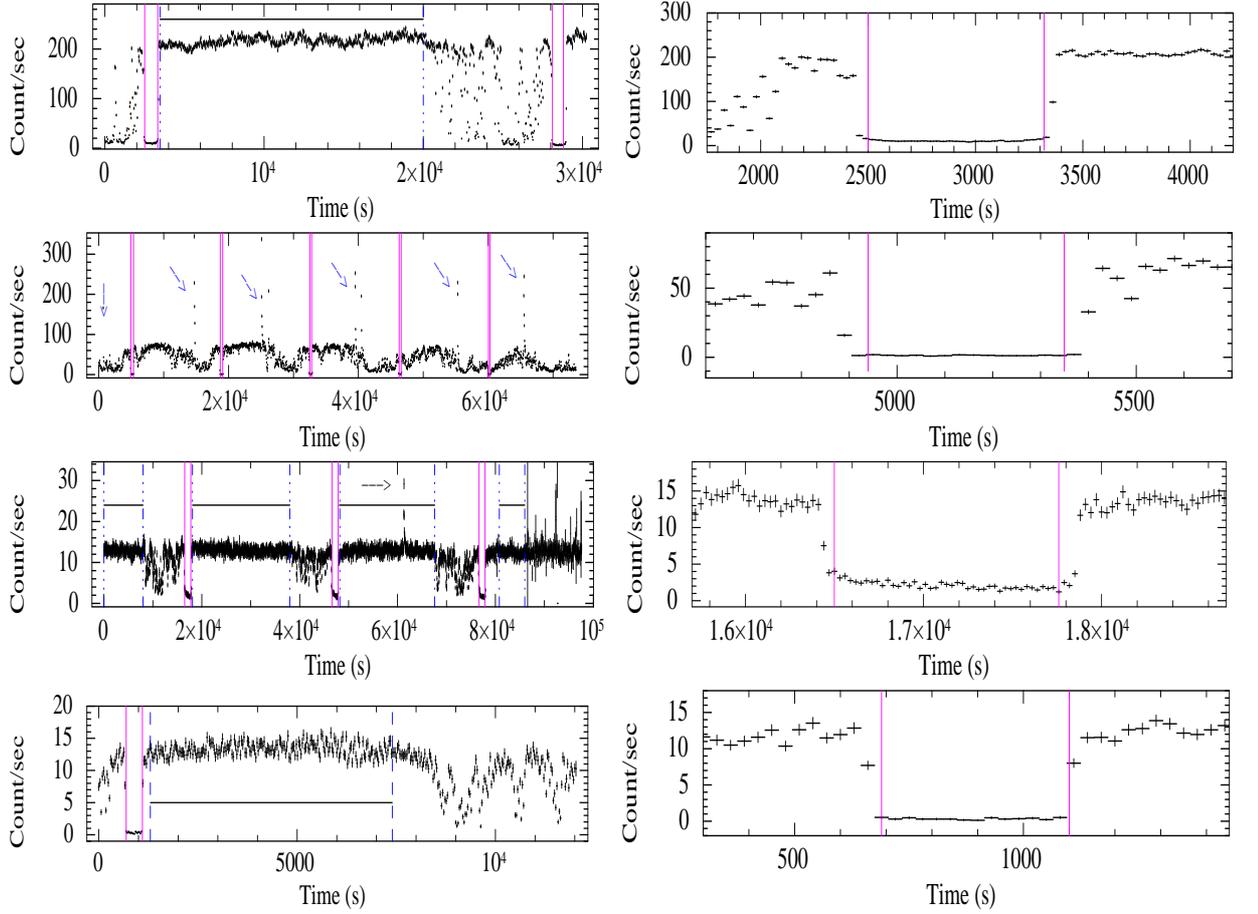

%  \centering
 \includegraphics[width=3cm,height=8.4cm, angle=-90]{vp_lcu_ecp_label_MXB_701_2_1.ps}
  \hspace{-0.45cm}
 \includegraphics[width=3cm,height=8.4cm,  angle=-90, trim={0.2cm, 0, 0, 0}, clip]{vp_zoom_lcu_ecp_label_MXB_701_1_gedit.ps}
%  \hspace{50cm}
 \includegraphics[width=3cm,height=8cm, angle=-90]{vp_lcu_label_ecp_OOE_EXO_ID401_16march.ps}
 \includegraphics[width=3cm,height=8.4cm, angle=-90]{vp_zoom_ecp1_16march18_2.ps}
  \includegraphics[width=3cm,height=8.4cm, angle=-90]{vp_lcu_AXJ_0402430401_ecp-OOE_26June19_1.ps}
 \hspace{-0.45cm}
 \includegraphics[width=3cm,height=8.3cm, angle=-90]{vp_lcu_label_ecp_AXJ_ID0402430401_2.ps}
 \includegraphics[width=3cm,height=8cm, angle=-90]{vp_lcu_ecp_OOE_label_XTE_1.ps}
 \hspace{-0.1cm}
 \includegraphics[width=3cm,height=8.4cm, angle=-90]{vp_zoom_lcu_ecp_label_XTE_2.ps}
  \captionsetup{width=20cm}
 \caption{The
eclipse and OOE events in the four lightcurves of the LMXBs;
one lightcurve is shown for each of the sources. The four lightcurves are of MXB 1659$-$298 (OB ID: 0008620701, top-most panels), EXO 0748$-$676 (OB ID: 0160760401, panels in the second row), AX J1745.6$-$2901 (OB ID: 0402430401, panels in the third row) and XTE J1710$-$281 (OB ID: 0206990401,   bottom-most panels).
 \textbf{Left figures (Full lightcurves):} The events within pink solid (blue dot-dashed/dashed) lines in the light-curves show eclipse (OOE) events.
 The  eclipse (OOE) spectrum has been obtained from these eclipse (OOE) events. Bursts (wherever found), dips and count rate 
 less than 70$\%$ of the peak count rate after excluding the bursts have been removed for the extraction of the OOE spectra. 
 \textbf{Right figures (Eclipse lightcurves):} Zoomed view of an eclipse event in the same lightcurves.  Similar extraction of the eclipse and OOE spectra has been done for other 
 observations of  each of the sources.  A single eclipse (OOE) spectrum has been obtained from multiple eclipse (OOE) events 
 available in an observation. Combined eclipse (OOE) spectrum of a source has been obtained by stacking all the  eclipse  
 (OOE) spectra of all the observations of a particular source.}
\label{mark_ecp_ooe_1}
\end{figure*}

 The total lightcurves of the sources show one or more eclipse events. 
 Figure \ref{mark_ecp_ooe_1} shows the eclipse and OOE events in the four lightcurves of the LMXBs,
 one lightcurve is shown for each of the source.  
 The Figure shows sharp fall of intensity from out-of-eclipse
 to eclipse in the four sources. This suggests that the direct emission from the neutron star, inner accretion disk, 
 boundary layer and corona is completely
 blocked during eclipse.
 \par
 We extracted two set of events:
 1) eclipse events 2) out-of-eclipse (OOE) persistent events from the 
 event files of each observation. For that we first identified the eclipse and OOE phases 
 from the background corrected source lightcurves
  of each observation. In each observation we noticed the start and end time of all the eclipse phases, 
  then using the  
  task {\small  GTIBUILD} we generated the good time intervals of all the
  eclipse phases covered in the observation. Then using the  
  task {\small  EVSELECT} we extracted the eclipse events for each observation. 
  To extract the persistent OOE events we removed the intervals of bursts and dips wherever found, and considered
  the intervals when the count rates are
  above 70$\%$ of the peak count rate. Then in a similar manner  using the above tasks 
  we extracted the persistent OOE events as we extracted the eclipse events. 
 \subsection{SPECTRAL ANALYSIS}
 For all the observations mentioned in Table \ref{mode}, we have extracted eclipse and OOE
 spectra. For both the eclipse and OOE phases, we extracted the spectra
from the single and double events 
(with PATTERN$\leq$4) excluding the events which are at the edge of a CCD
and at the edge to a bad pixel (with FLAG=0).  We  generated response and ancillary 
files using the 
task {\small  RMFGEN} and  {\small ARFGEN} respectively.
We  used the 
task {\small  EVSELECT}
to extract the spectra from the eclipse and OOE events using corresponding response and ancillary files.
We  rebinned the spectra with the 
task  {\small SPECGROUP} to have
at least 25 counts per bin to be able to
use  $\chi^{2}$ statistics.
\par
Each observation of AX J1745.6$-$2901 cover 2 to 4  eclipses, whereas that of  
EXO 0748$-$676  and MXB 1659$-$298 cover 4 to 8 eclipses
 and 1 to 2 eclipses respectively. The observation of XTE J1710$-$281 covers only one eclipse. 
 For each observation we extracted a single eclipse
spectrum, combining the events from all the eclipses present in that observation.
Similarly from each observation we extracted a single OOE spectrum, from the phases where the sources were
persistent. 
\subsubsection{\textbf{ECLIPSE AND OUT-OF-ECLIPSE SPECTRA OF THE LMXBs}}
We have fitted 
% % % %  
 eclipse  and OOE spectrum  of each observation of each source 
to see the time dependence of the characteristics of the medium surrounding the compact object. 
We have also obtained the composite eclipse and composite OOE spectra of 
MXB 1659$-$298, EXO 0748$-$676 and AX J1745.6$-$2901 from two, seven and seven observations respectively.
\par
 We have used {\small XSPEC} v12.11.0 \citep{Arnaud1996}
 for  the spectral fitting. The eclipse and OOE
spectra were primarily fitted with power-law ({\small XSPEC} model component \textit{powerlaw})
modified by photoelectric absorption due to gas, molecule and grain in the interstellar medium (ISM)
along the line of sight ({\small XSPEC} model component \textit{Tbabs}, \citealt{Wilms_McCray_00}).
Though we have extracted the source and background regions where excess emissions are not observed to be present, however there is a possibility that some photons are scattered into the line of sight confronting the surface of the dust grains present in the ISM.
So to correct for the effect of  
X-ray scattering by the dust grains in the ISM we used the {\small XSPEC} model component \textit{xscat} \citep{Smith_etal_2016}
and tied the value of interstellar hydrogen column density of this component with that (N$_{\rm H}$)
 associated with component \textit{Tbabs}.
The component \textit{xscat} incorporates the dependence of the scattering cross-section on energy,
 dust and the size of the 
extraction region. We have chosen the default dust model \citep{Mathis_etal_1977}
that comes with the \textit{xscat} code. The position of dust layer (Xpos) in a scale 0-1 is an input parameter in the component \textit{xscat}. Where 0
corresponds to the position of the observer and while 1 is at the source. The value 0.5 is default in the component. We have initially
kept it free while fitting the spectra. When its value could not be constrained then it is frozen to get best fits.
Looking at the value of $\chi^{2}$/DOF i.e. reduced $\chi^{2}$ ($\chi_{r}^{2}$) and the residuals in $\chi$ in the spectra,
we added other components  to obtain the best fit.
A few of the observations showed excess low energy 
emission, a  blackbody component was added to the model for these sources. 
{\small XSPEC} model component \textit{bbodyrad} and / or  \textit{diskbb} gave the best fit of these low energy excess emissions.
Excess narrow emission and / or absorption profiles were seen in a few spectra.
The excess and negative resisuals
at specific energies were fitted with Gaussian functions. {\small XSPEC} model component \textit{gaussian} and  \textit{gabs}
have been used for emission and absorption lines respectively. For some of the 
observations in the OOE phase an additional component for partial covering absorption by partially ionized material 
 ({\small XSPEC} model component \textit{zxipcf}) was required to fit the spectra. We have also stacked the eclipse (and OOE)
  spectra of each source obtained from different observations with the 
  task {\small EPICSPECCOMBINE}. Then changed the exposure of this stacked eclipse (and  OOE) spectrum of each 
  source to the total exposure of all the eclipse (and OOE) spectra avaiable
 for that particular source to obtain time averaged eclipse (and OOE) spectrum. In this way we obtained 
 a single combined eclipse (and OOE) spectrum of each source (except for XTE J1710$-$281,
 as this system has only one EPIC pn observation covering eclipse) just to see the average  variation of intensity with energy.
 We describe the details about the spectral fitting for all the sources in the following sections.
\\
\par
\textbf{MXB 1659$-$298:}\\
\\
% -------------------------------------------------
We have analyzed two EPIC pn observations of MXB 1659$-$298 (OB IDs: 0008620601, 0008620
701), covering 1-2 eclipses, 
carried out in an interval of $\sim$1 year. 
\citet{2001Sidoli_MXB}  have obtained the best fit of the persistent and eclipse
spectrum of the source with \textit{XMM-Newton} EPIC pn with a cutoff power-law, a blackbody emission, three Gaussian functions 
for two Fe absorption lines of 6.64 keV, 6.9 keV and 6.47 keV Fe emission line.
 They obtained the best fit of the eclipse spectrum with 
the same model. In the RGS persistent spectrum they detected absorption features from Oxygen and Neon ions with a 
wide range of ionization states. 
\cite{2006Trigo} have obtained best fit of the 
persistent spectrum of this source with \textit{XMM-Newton} EPIC pn with a power-law, a blackbody component modified with 
absorption due to both neutral and ionized absorbers and two Gaussian functions for 6.4 keV Fe K$_{\alpha}$ and 
0.99 keV emission line modified with
absorption from neutral  absorber. 
\par
 We first started fitting the eclipse spectrum of the observation of MXB 1659$-$298 with higher exposure time
 (OB ID: 008620701) between the two observations, we first checked the fitting  with a simple power-law modified with photoelectric absorption 
 and scattering in the ISM (Model \textit{Tbabs}$\times$\textit{xscat}$\times$\textit{powerlaw}) in the energy range of 0.3-10 keV.
 We get $\chi^{2}$ = 94.47 for 105 DOF and we accept this as the best fit. We obtain best fit of the eclipse spectrum
 of the other observation (OB ID: 0008620601) with the same model ($\chi^{2}$ = 21.75 for 25 DOF)
\par
 Fitting of the OOE spectrum with higher exposure time (OB ID: 008620701) with model
 \textit{Tbabs}$\times$\textit{xscat}$\times$\textit{powerlaw} shows
 some low energy excess in the residuals ($\chi^{2}$ = 381.04 for 110 DOF) in the energy range of
 3-10 keV. The low energy excess is well fitted with the addition of 
 component \textit{bbodyrad}, with $\chi^{2}$ = 186.89 for 108 DOF. Instead of blackbody component \textit{bbodyrad}
 if we use component \textit{diskbb}, the statistics is similar ($\chi^{2}$ = 186.28 for 108 DOF), but the disk
 normalization is low (N$_{\rm disk}$ = 0.44). Following \cite{Kubota_etal_98}
 and using this normalization value and distnace to the source \citep{2006Trigo},
 we get a very low value of projected
 real inner disk 
 radius of the accretion disk ($<$ standard radius of a neutron star), so we do not consider \textit{diskbb} 
 to fit the low energy excess.
 With the model \textit{Tbabs}$\times$\textit{xscat}(\textit{powerlaw} + \textit{bbodyrad})
 we see some negative residuals in the $6-7$ keV energy range. Addition of model component \textit{gabs}
 for the 6.75 keV absorption line improves the fit ($\chi^{2}$ =147.24 for 106 DOF).
% %   
 Addition of another \textit{gabs} component for
  6.99 keV absorption line further improves the fit ($\chi^{2}$ = 96.85 for 105 DOF). 
 We accept this as the best fit for the OOE spectrum of the source.  
 N$_{\rm H}$ could not be constrained and we freeze it to the value 
 following \citealt{2001Sidoli_MXB} (Table \ref{phenomeno}).
 In a similar manner we approached to fit the OOE spectrum of the other observation (OB ID: 008620601).
 We obtained best fit with the model 
 \textit{Tbabs}$\times$\textit{xscat}$\times$\textit{gabs}(\textit{powerlaw} + \textit{bbodyrad}) with 
 $\chi^{2}$ = 142.21 for 162 DOF. Addition of the emission lines in the two OOE spectra are significant
 at a confidence level $\textgreater$ 99.99$\%$.
\par
\begin{figure*}
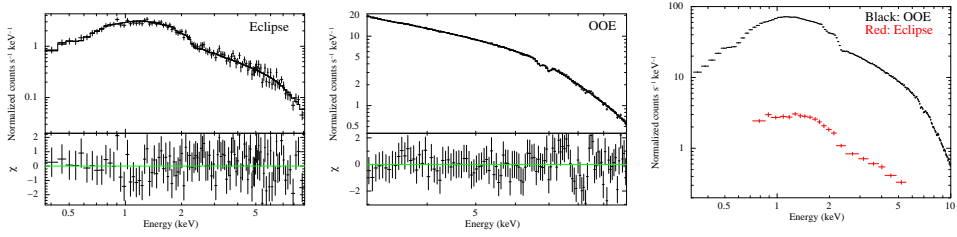

 \centering
 \includegraphics[scale=0.16, angle=-90]{xscat_mxb_ecp_id701_6_5sept-21_1.ps}
\includegraphics[scale=0.16, angle=-90]{xscat_mxb_ooe_id701_4_frm-3kev_4_2.ps}
\includegraphics[scale=0.16, angle=-90]{MXB_both_1dec20_2.ps}
 \caption{Left and middle figure show best fit of the eclipse and OOE spectrum of MXB 1659$-$298 respectively having highest exposure time between the two soft-state observations (OB ID: 0008620701). The bottom panels in these two figures are the contribution of $\chi$ towards each bin for the best fit. The right figure shows both time averaged combined eclipse (red) and OOE (black) spectra of MXB 1659$-$298 obtained from two observations.}
 \label{MXB_specs}
 \end{figure*}
\begin{figure*}
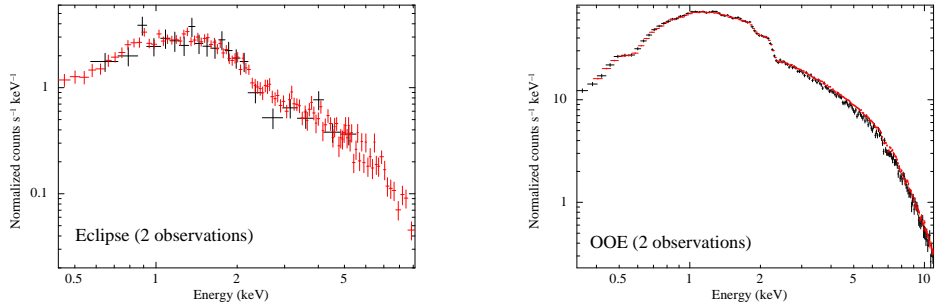

\centering
\includegraphics[scale=0.22, angle=-90]{composite_MXB_ecp_2.ps}
\hspace{1cm}
\includegraphics[scale=0.22, angle=-90]{composite_MXB_ooe_2.ps}
\caption{Composite eclipse spectra (left)
and OOE spectra (right) of MXB 1659$-$298 for the two EPIC-pn observations, observed in soft-state.}
\label{MXB_comp}
\end{figure*}

  N$_{\rm H}$  and $\Gamma$ during eclipse (OOE) phase 
 have been found in the range of 25-26$\times$10$^{20}$ cm$^{-2}$ (28-35$\times$10$^{20}$ cm$^{-2}$)
 and 1.52-1.57 (1.75-2.07) respectively. The 3-10 keV flux obtained from both OOE  (and eclipse) 
 spectra are quite similar.  The OOE to eclipse flux ratio obtained in the energy range of ($3-10$) keV has been found to be $21-24$ 
 for the two observations. To be consistent with the obsevations of other three sources reported in this text 
 we have chosen the lower energy to be 3 keV for the flux estimation. The position of the dust layer (Xpos) is 
 obtained as 0.98$^{+0.01}_{-0.05}$ for the OOE spectrum of OB ID 008620601. In other eclipse and OOE spectra
 of the source its value could not be constrained and it is frozen with the value 0.98.
 The details of the best fit models and corresponding spectral parameters are given in Table \ref{models}, \ref{phenomeno}, \ref{phenomeno_line} 
 respectively. Figure \ref{MXB_specs} shows  best fit of the eclipse and OOE spectrum of MXB 1659$-$298 for observation which has higher
 exposure time (OB ID: 0008620701) between the two observations in the left and middle figure respectively. The right figure shows the 
 time averaged combined OOE (black) and eclipse (red) spectrum of the source obtained from the two observations.
  Figure \ref{MXB_comp} shows the composite eclipse (left figure)
 and OOE spectra  (right figure) respectively for the two observations, where both eclipse (OOE) spectra are observed to almost overlap.
% =============
   \\
% -------------------------------------------------
\par
\textbf{EXO 0748$-$676:}\\
\\
% -------------------------------------------------
We  have analyzed  seven EPIC pn observations of EXO 0748$-$676 carried out over a period of $\sim$2 months. 
Each observation covers minimum 4 to maximum 8 eclipses. 
Triplets of thermonuclear X-ray bursts \citep{2007Boirin}, X-ray burst reprocessing \citep{2012Paul},
 orbit to orbit variations of the X-ray dips and orbital modulations of the optical light curves 
\citep{2017Gayathri} have been reported from these observations. \citealt{2017Gayathri} reported the source to be in the hard state during these seven observations.
\cite{2008Cottam} reported 
gravitationally redshifted absorption lines in the 28 burst spectra of this source with \textit{XMM-Newton} RGS.
 \citet{2017Gayathri} have obtained the best fit of the persistent \textit{XMM-Newton} spectra with a power-law modified with neutral 
 and parital covering ionized absorption along with two Gaussian functions
 for 0.569 keV (fixed) and 0.915 (fixed) keV emission lines modified with photoelectric absorption along the
 line of sight due to neutral absorber.
%  *
Increase of covering fraction of the ionized absorber and a decrease 
in the ionization parameter has been observed
 from persistent emission to dipping (\citealt{2006Trigo}, \citealt{2017Gayathri}).  
%  
% *************************************
\par
 We first checked the fitting of the eclipse spectrum of EXO 0748$-$676 which has highest exposure
  time amongst the seven observations (OB ID: 0160760301). 
 The model  \textit{Tbabs}$\times$\textit{xscat}$\times$
 \textit{powerlaw} gives a $\chi^{2}$ = 245.38 for 88 DOF  
 in the energy range of 0.3-10 keV. We 
 see some low energy excess and use the component \textit{bbodyrad} to account for it, the fit is improved with
 $\chi^{2}$ = 112.96 for 86 DOF. Since during eclipse the inner accretion disk is expected to be 
 blocked by the companion, so we
 use blackbody component \textit{bbodyrad} instead of component \textit{diskbb}.
 We see an excess $\sim$0.5 keV, addition of an emission line of 0.55 keV
 with model component \textit{gaussian} significantly
 improves the fit ($\chi^{2}$ = 97.58 for 84 DOF),  so we report model
 \textit{Tbabs}$\times$\textit{xscat}(\textit{powerlaw} + \textit{bbodyrad} + \textit{gaussian}) 
 as the best fit model for this eclipse spectrum. In a
 similar manner we fitted the other eclipse spectra of this source. Other six observations required $1-5$
 emission lines for the best fits. For example the observation with OB ID: 
 0160760201 required five emission lines for the best fit. Fitting this spectrum with 
 the modeel \textit{Tbabs}$\times$\textit{xscat}(\textit{powerlaw} + \textit{bbodyrad}),
 we get a $\chi^{2}$ = 191.98 for 66 DOF with
 some positive residuals at specific energies below 2 keV and near 6.4 keV. When we 
 add component \textit{gaussian} for these excess, the fit gradually improves. For example
 adding this component (\textit{gaussian}) for emission lines at 0.54 keV, 0.89 keV, 1.39 keV, 1.80 keV
 and 6.44 keV we get $\chi^{2}$ of 109.64, 79.23, 71.60, 64.20 and 56.51 for 63, 
 61, 59, 57 and 55 DOF respectively. In the similar approach we
 obtain the best fit of the other eclipse spectra of this source. Addition of the 
 lines are significant at a confidence
 level $>$99.99$\%$. Except for OB ID 0160760101, N$_{\rm H}$ could not be constrained and it is frozen with 
 3$\times$10$^{20}$ cm$^{-2}$ to obtain best fit of the eclipse spetra.
 \par
 Fitting of the OOE spectrum of OB ID: 0160760301 having highest exposure amongst the seven OOE observations
 with the simple model  \textit{Tbabs}$\times$\textit{xscat}$\times$\textit{powerlaw} 
 gives $\chi^{2}$ = 90417 for 176 DOF in the energy range of 0.3-10 keV. We observe some  
 low energy excess and positive residuals in $5-6$ keV energy range. We add 
 the blackbody component
 \textit{diskbb} to account for the low energy excess, the fit is improved with $\chi^{2}$ = 2269 for 174 DOF. 
 Adding 0.59 keV emission line with
 component \textit{gaussian} makes the fit better ($\chi^{2}$ = 1269 for 171 DOF). We observe some continuous negative residuals
 at low energies, to 
 account for it we add a partial covering absorption model by partially ionized material (component \textit{zxipcf}) following
 \cite{2017Gayathri}.
 The fit is significantly improved with $\chi^{2}$ = 142.80 for 168 DOF. Instead of ionised absorption component (\textit{zxipcf}),
 if we use partial covering absorption model by neutral material (component \textit{pcfabs}), the $\chi^{2}$ is 187.81 for 169 DOF.
 For this reason we use the component \textit{zxipcf} instead of \textit{pcfabs}.  So we accept 
 the model  $\textit{Tbabs}$$\times$\textit{xscat}$\times$\textit{zxipcf}(\textit{powerlaw} + \textit{diskbb}) + \textit{Tbabs}$\times$\textit{xscat}$\times$\textit{gaussian} as the best fit model for this eclipse spectrum. We use \textit{diskbb} component to account for the blackbody radiation for the OOE spectra, as the inner disk is 
 visible when the compact object is out of eclipse. We have obtained the best fit of the OOE spectra of other observations of this source 
 in a similar manner.
 In the three observations we required $\sim$6.8 keV absorption line (OB ID: 0160760401, 0160760601 and 0160760801). Addition of the 
 emission line is significant at a confidence level $>$99.99$\%$. For the best fit of the 
 two observations (OB ID: 0160760401 and 0160760801) no emission line is observed.  In one observation (OB ID: 0160760101)
 we have observed a significant positive residual at low energies even after fiting with the model 
 $\textit{Tbabs}$$\times$\textit{xscat}$\times$\textit{zxipcf}$\times$\textit{gabs}(\textit{powerlaw} + \textit{diskbb}) + \textit{Tbabs}$\times$\textit{xscat}$\times$\textit{gaussian} ($\chi^{2}$ 400.14 and 168 DOF). Adding a blackbody component \textit{bbodyrad}
 significantly improves the fit ($\chi^{2}$ 195.87 for 166 DOF), we accept this fit as the best fit 
 for this observation during OOE phase ($\textit{Tbabs}$$\times$\textit{xscat}$\times$\textit{zxipcf}$\times$\textit{gabs}(\textit{powerlaw} + \textit{diskbb} + \textit{bbodyrad}) + \textit{Tbabs}$\times$\textit{xscat}$\times$\textit{gaussian}).
 \par
\begin{figure*}
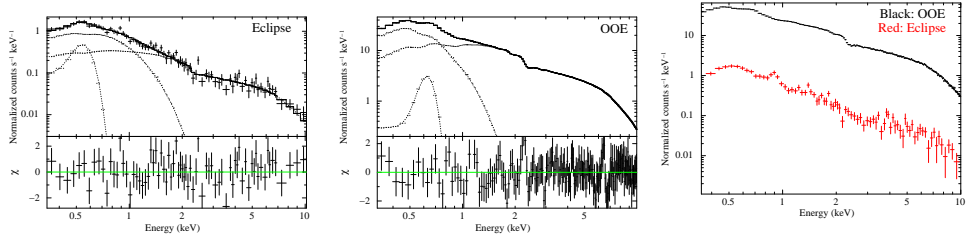

 \centering
\includegraphics[scale=0.16, angle=-90]{xscat_exo_ecp_id301_7.ps}
\includegraphics[scale=0.16, angle=-90]{xscat_exo_ooe_id301_2.ps}
\includegraphics[scale=0.16, angle=-90]{both_emrgn_EXO_2.ps}
\caption{Eclipse (left figure) and OOE (middle figure) spectrum of EXO 0748$-$676  respectively with highest exposure time amongst the seven hard-state  observations (OB ID: 0160760301). The bottom panels in these two figures are the contribution of $\chi$ towards each bin for the best fit.
The right figure shows both time averaged combined eclipse (red) and OOE (black) spectra of EXO 0748$-$676 obtained from seven observations.}
 \label{EXO_specs}
 \end{figure*}
\begin{figure*}
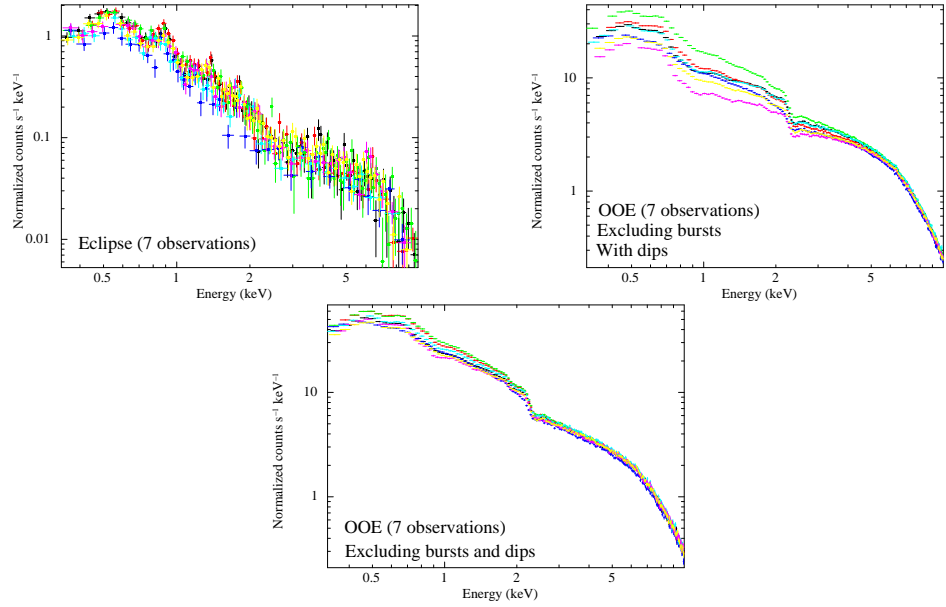

\centering
\includegraphics[scale=0.22, angle=-90]{composite_7_EXO_ecp_2.ps}
\hspace{1cm}
\includegraphics[scale=0.22, angle=-90]{seven_OOE_with_dips-but-burst-free_2.ps}
\includegraphics[scale=0.22, angle=-90]{seven_OOE_persistant_dips-n-burst-free_2.ps}
\caption[Composite eclipse and OOE spectra of EXO 0748$-$676]{Composite eclipse spectra (top left), OOE spectra including dips intervals (top right), OOE spectra excluding the dips intervals (bottom panel) of EXO 0748$-$676 for the seven EPIC-pn observations, observed in hard-state.}
\label{EXO_comp}
\end{figure*}

 The Temperature associated with component \textit{bbodyrad} is high (4.16$^{+0.72}_{-0.44}$ keV) with a small value of normalization(0.08$^{+0.03}_{-0.03}$).
% % %  
% % %  
 N$_{\rm H}$  and $\Gamma$ during eclipse (OOE) phase 
 have been found in the range of 3-4 $\times$10$^{20}$ cm$^{-2}$ (11 $\times$10$^{20}$ cm$^{-2}$ frozen) and
 0.87-1.16 (1.27-2.24) respectively. The (3-10) keV flux values are quite similar
  in all the observations during eclipse. The OOE to eclipse flux ratios have been
  found  in the range of  $\sim$(43-54) for the seven observations. The ionized absorbers have been found to cover (55-92)$\%$ 
  (as CvF, the covering factor $\sim$0.55-0.92) of the source emission during out-of-eclipse. 
  The position of the dust layer (Xpos) could not be constrained and is frozen at 0.98 to obtain the best 
  fit of the eclipse and OOE spectra. The details of the best fit model and spectral 
 parameters are given in Table \ref{models}, \ref{phenomeno}, \ref{phenomeno_line} and  \ref{EXO_OOE_zxipcf} respectively. Figure \ref{EXO_specs} shows  best fit of
 the eclipse (left) and OOE (middle) spectrum of EXO 0748$-$676 for observation 
 which has highest exposure time among the seven observations. The right panel of this figure shows
 the time averaged combined OOE (black) and eclipse (red) spectra obtained from combinig the seven OOE and eclipse observations respectively.
 Figure \ref{EXO_comp} shows
 the composite eclipse (top left) and OOE (top right 
 and bottom) spectra of the seven observations. The OOE spectra obtained including 
 and excluding the dips intervals
 are shown in the top right panel  and bottom panel respectively.\\
 \\
%  
% **************************
% % % % 
\par
\begin{figure*}
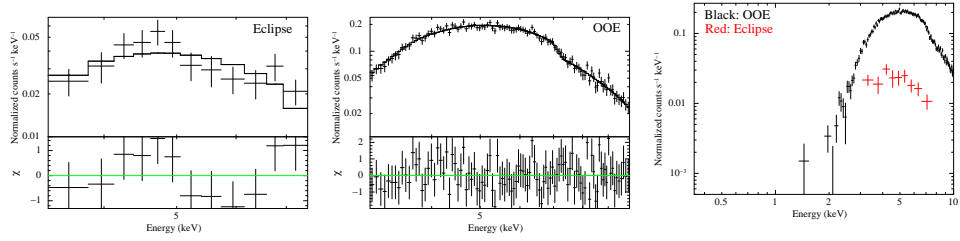

 \centering
\includegraphics[scale=0.16, angle=-90]{xscat_axj_hard_ecp_id-0723410501_2.ps}
\includegraphics[scale=0.16, angle=-90]{xscat_axj_hard_ooe_id-0723410501_2.ps}
\includegraphics[scale=0.16, angle=-90]{AXJ-hard_both_1dec20_2.ps}
\caption{Eclipse (left figure) spectrum of AX J1745.6$-$2901 (OB ID: 0723410501) with highest exposure time amongst the three hard-state  observations and corresponding OOE (middle figure) spectrum. The bottom panels in these two figures are the contribution of $\chi$
    towards each bin for the best fit. The right figure shows both time averaged combined eclipse (red) and OOE (black) spectra of AX J1745.6$-$2901 obtained from three hard-state observations.}
    \label{AXJ_hard_spec}
     \end{figure*}

\textbf{AX J1745.6$-$2901:}\\
\\
% -------------------------------------------------
 We  have analyzed  seven EPIC pn observations of AX J1745.6$-$2901 carried out over a period of $\sim$8 and half years. 
 Each observation covers minimum 2 to maximum 4 eclipses. We first checked the fitting of the eclipse spectrum of the
 hard state observation of the source which has the highest exposure time (OB ID: 0723410501) during 
 eclipse in the energy range of 2-10 keV, as below
 2 keV the photon counts are almost nil. The best fit was
 obtained with simple model \textit{Tbabs}$\times$\textit{xscat}$\times$\textit{powerlaw} with $\chi^{2}$ = 10.74, for 10 DOF.
 In a similar manner the best fit of the eclipse spectra of other hard-state observations were obtained
 with the same model stated above. The OOE spectrum of the observation (OB ID: 0723410501) 
 were also best fitted with the simple model (\textit{Tbabs}$\times$\textit{xscat}$\times$\textit{powerlaw}) with $\chi^{2}$ = 
 74.65 for 99 DOF. Best fit of the OOE spectrum of the other two observations were also obtained with this model.  
 N$_{\rm H}$  and $\Gamma$ during eclipse (OOE) phase have been found in the
 range of ($9.39-19.50$) $\times$10$^{22}$ cm$^{-2}$ (($17.15-19.50$) $\times$10$^{22}$ cm$^{-2}$) 
 and $2.05-2.65$ ($1.65-1.89$) respectively. \cite{2015ponti_Fe} reported an absorbed power-law model 
 for the best fit of the hard state \textit{XMM-Newton} persistent spectra of the source.
 The details of the best fit model and spectral parameters are given in Table \ref{models}, \ref{phenomeno}, \ref{phenomeno_line}
 respectively. The left panel of Figure \ref{AXJ_hard_spec} shows  best fit of the eclipse spectrum of AX J1745.6$-$2901 
 for observation which has highest exposure time during eclipse among the three hard state observations, 
 the middle panel shows corresponding OOE spectrum. The right panel shows the time averaged combined OOE (black) 
 and eclipse (red) spectrum of the 
 source in hard state obtained from the three OOE and eclipse observations respectively.
 \par
 The eclipse spectrum of the soft state observation (OB ID: 0762250301) of AX J1745.6$-$2901 with highest exposure time was first fitted with model 
 \textit{Tbabs}$\times$\textit{xscat}$\times$\textit{powerlaw} in the energy range of 2-10 keV, the fit gives a $\chi^{2}$ 42.53 for DOF 50. We accept this
 as the best fit. For the other two observations also best fit of the eclipse spectra are obtained with the same model. The OOE spectrum of the 
 observation (OB ID: 0762250301) was first fitted with the simple model (\textit{Tbabs}$\times$\textit{xscat}$\times$\textit{powerlaw})
  in the energy range of 0.3-10 keV,
 a $\chi^{2}$ 383.95 for DOF 110 was observed. Some low energy excess and both negative and positive residuals are observed
 in 6-7 keV energy range. Addition of blackbody component \textit{bbodyrad} improves the fit ($\chi^{2}$ 269.08 for DOF 108).
 Instead of \textit{bbodyrad} if we use \textit{diskbb} component then the disk normalization is very low (0.43).
 Following \cite{Kubota_etal_98}
 and from this normalization and distance to the source, the value of
 the apparent inner disk radius is estimated to be significantly smaller than the radius of the netron star. So we accept component
 \textit{bbodyrad} to fit the low energy excess in this spectrum. 
  Adding 6.71 keV and 6.96 keV absorption lines we get better fits ($\chi^{2}$  
  220.43 and 139.63 for 107 and 106 DOF  respectively). Addition of 
 an emission line of 6.4 keV finally gives the best fit 
 (\textit{Tbabs}$\times$\textit{xscat}$\times$\textit{gabs$_{1}$}$\times$\textit{gabs$_{2}$}(\textit{powerlaw} + \textit{bbodyrad} + \textit{gaussian})) 
 with $\chi^{2}$  132.08 for 105 DOF. The addition of the absorption and emission lines is significant at a confidence level $>$99.99$\%$.
 \par
\begin{figure*}
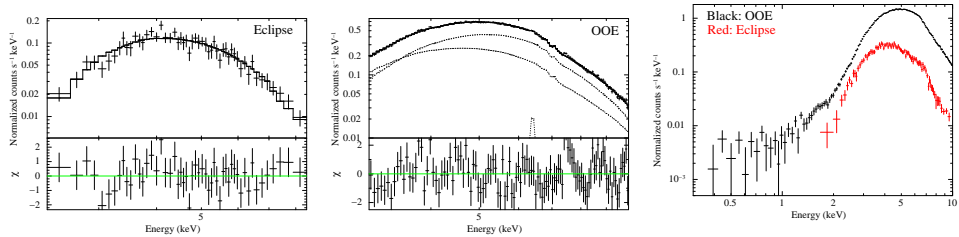

 \centering  
\includegraphics[scale=0.16, angle=-90]{xscat_axj_soft_ecp_0762250301_2.ps}
\includegraphics[scale=0.16, angle=-90]{xscat_axj_soft_ooe_0762250301_bbrd_good_6.ps}
\includegraphics[scale=0.16, angle=-90]{AXJ-soft_both_1dec20_2.ps}
    \caption{Eclipse (left figure) and OOE (middle figure) spectrum of AX J1745.6$-$2901 (OB ID: 0762250301) with highest exposure time amongst the four soft-state  observations. The bottom panels in these two figures are the contribution of $\chi$
    towards each bin for the best fit. The right figure shows both time averaged combined eclipse (red) and OOE (black) spectra of AX J1745.6$-$2901 obtained from four soft-state observations.}
    \label{AXJ_soft_spec}
 \end{figure*}

 In a similar manner we fitted
 the OOE spectra of the other three soft state observations. The best fit of the two OOE spectra (OB ID: 0402430401, 0724210201)
 were obtained with the same model except the emission line. The best fit of the OOE spectrum of the other observation 
 (OB ID: 0402430301) was obtained with the model
 \textit{Tbabs}$\times$\textit{xscat}$\times$\textit{gabs$_{1}$}$\times$\textit{gabs$_{2}$}$\times$\textit{powerlaw}. 
 N$_{\rm H}$  and $\Gamma$ during eclipse (OOE) phase  have been found in the range of (17.14-19.23) $\times$10$^{22}$ cm$^{-2}$
 (($16.35-20.58$) $\times$10$^{22}$ cm$^{-2}$) and  $3.03-3.83$ ($1.65-2.36$) respectively. $\Gamma$ is higher during eclipse phases compared to OOE
 in all the hard and soft sttate observations of AX J1745.6$-$2901. 
 The OOE to eclipse flux ratio in the energy range of (3-10) keV is in the range of $\sim$(6-10) for the source. 
 We have detected Fe XXV and Fe XXVI absorption lines in the soft states but did not find any absorption feature in the hard states
 (upper limits of strength of Fe  XXV and Fe XXVI absorption lines in hard state observations are 0). 
 \cite{2015ponti_Fe} also has observed these absorption lines only in soft states in this source. 
 The position of the dust layer (Xpos) could not be constrained and is frozen at 0.98 to obtain the best 
  fit of the eclipse and OOE spectra of the source in both soft and hard states.
 The best fit model and the details of the best fit spectral parameters of the combined eclipse and OOE spectra of AX J1745.6$-$2901
  are given in Table \ref{models}, \ref{phenomeno} and \ref{phenomeno_line} respectively.
  Figure \ref{AXJ_soft_spec} and \ref{AXJ_hard_spec} show the eclipse (left figures) and OOE spectra 
  (middle figures)  of AX J1745.6$-$2901 for the soft and hard state observations respectively. The right figures show
  the time averaged combined OOE (black) and eclipse (red) spectrum of the source obtained from the 
  three hard state and four soft state
  observations respectively.
 \par
\begin{figure*}
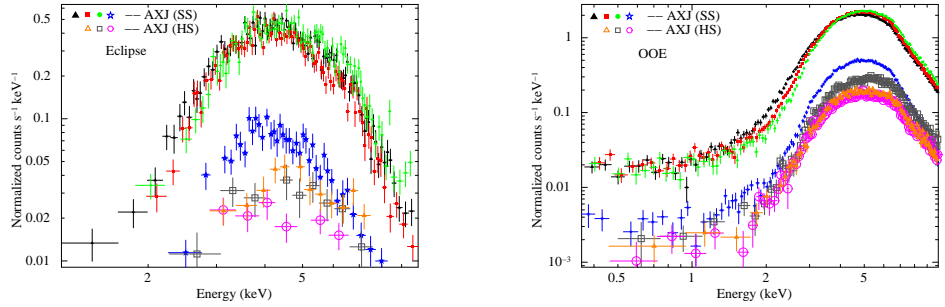

\centering
\includegraphics[scale=0.22, angle=-90]{composite_all_AXJ_ecp_4.ps}
\hspace{1cm}
\includegraphics[scale=0.22, angle=-90]{composite_all_AXJ_OOE_4.ps}
\caption{Composite eclipse (left) and OOE (right) spectra  of AX J1745.6$-$2901 
for seven EPIC-pn observations. The soft-state spectra are represented with solid triangle
(black), box (red), circle (green) and star (blue) marker and the hard-state spectra are 
shown with hollow box (grey), circle (magenta) and triangle (orange) marker in both the 
figures. The intensity of the soft-state spectra are higher compared to that of the hard-state
spectra in both phases. Intensity of a soft-state OOE spectrum (right figure, blue solid star) is  more than five times lower than that of the
three  other soft-state OOE spectra. Corresponding to this observation, the intensity drops by similar factor in the reprocessed
eclipse spectrum (left figure, blue solid star) also.}
\label{AXJ_all_comp}
\end{figure*}

Figure \ref{AXJ_all_comp} shows the composite eclipse  (left figure) and out-of-eclipse  (right figure)  
spectra respectively for the 
  soft  and  hard state observations of AX J1745.6$-$2901.  
 The intensity of the four soft-state eclipse (OOE) spectra (black, red, green, blue)
 are seen to be higher
 compared to three hard-state eclipse (OOE) spectra (orange, grey, magenta).  Intensity of a soft-state OOE spectrum (blue solid star, right figure) 
 is  more than 5 times lower than three other
soft-state OOE spectra. Corresponding to this observation, the intensity drops by similar factor in the reprocessed 
eclipse spectrum (left figure) also.\\
 \\
 \\
\textbf{XTE J1710$-$281:}\\
\\
 We have analyzed one EPIC pn observation of XTE J1710$-$281, which covers single eclipse. 
 \citet{2009Younes} reported the \textit{XMM-Newton} persistent and dip spectrum fitting
 with a simple power-law modified for line of sight absorption due to neutral materials along with a partial covering
 abosrption due to neutral absorber.
 They obtained best fit of these spectra
 when they replace the partial covering abosrption due to neutral absorber with that due to ionized absorber. 
 The parameters of the ionized absorber could not be constrained for the persistent spectrum. They have not obtained 
 any signature of highly ionized Fe lines.
\citet{2018raman} obtained best fit of the time average \textit{Chandra} spectrum with a simple
power-law  modified with line of sight absorption due to neutral material and partially covered
ionized materials. To get the best fit of the \textit{Suzaku} spectrum they required three Gaussian functions, 
two for 6.6 keV and 7.01 keV absorption lines and one for 0.72 keV emission line. Highly ionized Fe
line has been obtained for the first time in the source (\citealt{2018raman}). They have also obtained intensity 
resolved spectral fit with \textit{Chandra} and \textit{Suzaku}, and observed an increase of column density of the 
ionized absorber and a decrease of ionization state from the persistent to dip spectra.
\par
\begin{figure*}
\centering
   \includegraphics[scale=0.22, angle=-90]{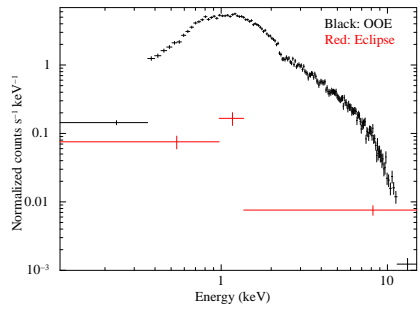}
\caption{Eclipse (red) and OOE (black) spectra of XTE 1710$-$281}
\label{XTE_spec}
\end{figure*}

The eclipse spectrum has very limited statistics leading to only a few bins in the binned spectrum.  
We fitted the OOE spectrum first with a simple \textit{powerlaw} component modified with the
line of photoelectric absorption and scattering in the ISM
(\textit{Tbabs}$\times$\textit{xscat}$\times$\textit{powerlaw}, $\chi^{2}$ = 197.74 for 150 DOF). 
When we add partial covering absorption due to neutral absorber or that due to ionized absorber,
the fit does not improve ($\chi^{2}$ of 198.32 and 195.51 for 148  and 147 DOF respectively).
So we accept the simple model stated above as the best fit model.
Only for the estimation of the flux ratio between the OOE and eclipse phases
we fitted the eclipse spectrum with model \textit{Tbabs}$\times$\textit{xscat}$\times$\textit{powerlaw} 
and the OOE to eclipse flux ratio has been found to be $\sim$129. The position of the dust layer (Xpos) could not
be constrained and is frozen at 0.98 to obtain the best fit of the spectra.
The details of the spectral parameters 
of the OOE spectrum of this source are given in Table \ref{phenomeno}.
The N$_{\rm H}$ is similar
to that obtained in the two observations of  MXB 1659$-$298.
Figure \ref{XTE_spec} shows both eclipse and OOE spectra of XTE J1710$-$281 in one panel. \\
\\
\\
\par
\subsection{\textbf{To investigate  the effect of dust scattered emission in the eclipse (OOE) spectra}}
To check the effect of dust scattering in the eclipse (OOE) spectra we have fitted all the 
spectra including and excluding the dust scattering component \textit{xscat} and compared the spectral parameters. We have not observed any change.
To investigate further we have also obtained the low energy X-ray flux 
in 0.3-3 keV energy range including and excluding the component \textit{xscat}, since the dust scattering of X-ray photons
mostly affect the photon count distribution with respect to energy in low energy X-ray regime,
as the dust scattering cross section $\propto$ E$^{-2}$ \citep{Smith_etal_2016}. We obtain same flux
including and excluding the component \textit{xscat} in 0.3-3 keV energy range (Table \ref{compare_xscat_NOxscat}). This also confirms that the emission received during
eclipses are free from dust scattered emission and are from the binary system only.
\par
\begin{figure*}
\centering
\includegraphics[scale=0.46, angle=-90]{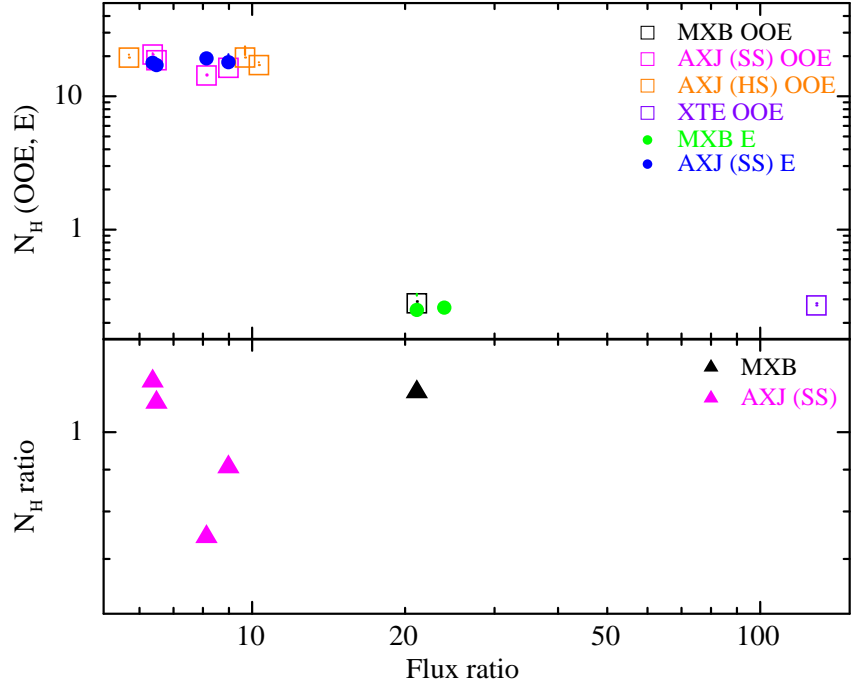}
\caption{N$_{\rm H}$ during OOE and eclipse (top panel) and their ratio (bottom panel) plotted with OOE to eclipse
 flux ratio for the three eclipsing LMXBs. In the top panel black, pink, orange and purple boxes denote OOE N$_{\rm H}$ 
 of MXB 1659$-$298, AX J1745.6$-$2901 soft state, AX J1745.6$-$2901 hard state and XTE J1710$-$281 observations respectively, 
 whereas green and blue circles represent N$_{\rm H}$ during eclipse of MXB 1659$-$298 and AX J1745.6$-$2901 
 soft state observations respectively. The black and pink triangles in the bottom panel represent OOE to 
 eclipse N$_{\rm H}$ ratio for MXB 1659$-$298 and AX J1745.6$-$2901 soft state observations respectively. The
 observations for which  N$_{\rm H}$ could not be constrained are not plotted here.}
\label{flux-rato_vs_NH}
\end{figure*}

% % % % To check further the presence of dust along and /or close to the line of sight, we have plotted the line of sight column density (N$_{\rm H}$) with eclipse to OOE flux ratio (X-ray reprocessing efficiency) for all the observations (Figure \ref{fluxratioNH}) except that of XTE J1710$-$281, as for this source the statistics of 
% % % % the eclipse spectrum is poor. In Figure \ref{flux-rato_vs_NH} we have plotted N$_{\rm H}$ during OOE and eclipse phase with OOE to eclipse flux ratio.
% 
To check further the presence of dust along and /or close to the line of sight, we have plotted the line of sight column density (N$_{\rm H}$) with OOE to eclipse flux ratio for all the observations (Figure \ref{flux-rato_vs_NH}) except that of XTE J1710$-$281, as for this source the statistics of the eclipse spectrum is poor. 
We do not observe correlation between the N$_{\rm H}$ and the flux ratio, which indicates negligible effect of dust scattered photons from a region along and /or close to the line of sight.
In the source EXO 0748$-$676, thermonuclear X-ray bursts have been observed during eclipses
(\citealt{1986Parmar_etalPpaper1}, \citealt{1985Parmar}, \citealt{Wolff_etal_2009})
which provides an evidence that the emission observed during
eclipses are the reflection from a region which is close to the neutron star. Otherwise if the emission during eclipses were due to the reflection from the
dust distributed in a larger region, then the bursts would be delayed by different amount from different dust layers, hence would not be detected. There is 
also an indication of detection of thermonuclear X-ray bursts in MXB 1659$-$298 during eclipse \citep{Wijnands_etal_2002}, 
but to be consistent they have analyzed the bursts outside eclipse and dips in that work.
% ========================================================================================
% ========================================================================================
% 
% +++++++++++++++++++++++++++
\begin{table*}
\footnotesize
 \label{table}
 \caption{The EPIC pn observation log of four eclipsing LMXB systems. All the observations were carried out in IMAGING mode.}
 \centering
% \resizebox{1.05\linewidth}{!}
 \begin{tabular}{cccccc}
\hline
Source 	   		& Observation ID & Date	of 		     & Submode 	      & Spectral 	& No. of \\
			&		 & observation  			 &		      & state		& eclipses\\
 \hline
 &&&&& \\
 MXB 1659$-$298		&0008620601      & 2000-03-22	 	       &  PrimeSmallWindow  &Soft    		& 1				\\
 MXB 1659$-$298		&0008620701      & 2001-02-20	 	       &  PrimeSmallWindow  &Soft  		& 2				\\
 \hline
 &&&&& \\
 EXO 0748$-$676          &0160760101      & 2003-09-19	 	       &  PrimeSmallWindow  &Faint-hard$^{c}$ & 7     		\\
 EXO 0748$-$676          &0160760201      & 2003-09-21	 	       &  PrimeSmallWindow  &Faint-hard$^{c}$ & 7    		\\
 EXO 0748$-$676          &0160760301      & 2003-09-23	 	       &  PrimeSmallWindow  &Faint-hard$^{c}$ & 8     		\\
 EXO 0748$-$676          &0160760401      & 2003-09-25	 	       & PrimeSmallWindow   &Faint-hard$^{c}$ & 5	   		\\
 EXO 0748$-$676          &0160760601      & 2003-10-21	 	       &  PrimeSmallWindow  &Faint-hard$^{c}$ & 4    		\\
 EXO 0748$-$676          &0160760801      & 2003-10-25	 	       &  PrimeSmallWindow  &Faint-hard$^{c}$ & 4   		\\
 EXO 0748$-$676          &0160761301      & 2003-11-12	 	       &  PrimeSmallWindow  &Faint-hard$^{c}$ & 7   		\\
%  &&&&& \\
  \hline
   &&&&& \\
 AX J1745.6$-$2901	& 0402430301     & 2007-04-01		       &PrimeFullWindow     & Soft$^{a}$ 	& 3			\\
 AX J1745.6$-$2901	& 0402430401     & 2007-04-03		       &PrimeFullWindow     & Soft$^{a}$ 	& 3			\\
 AX J1745.6$-$2901	& 0723410301     & 2014-02-28		       &PrimeFullWindow     & Hard$^{b}$ 	& 2			\\
 AX J1745.6$-$2901	& 0723410401     & 2014-03-10		       &PrimeFullWindow     & Hard$^{b}$ 	& 2			\\
 AX J1745.6$-$2901	& 0723410501     & 2014-04-02		       &PrimeFullWindow     & Hard$^{b}$ 	& 2			\\
 AX J1745.6$-$2901	& 0724210201     & 2013-08-30 		       &PrimeFullWindow     & Soft$^{a}$ 	& 2			\\
 AX J1745.6$-$2901	& 0762250301     & 2015-09-27	 	       &PrimeFullWindow     & Soft		& 4			\\				
%   &&&&& \\
  \hline
  &&&&& \\
   XTE J1710$-$281         &0206990401     & 2004-02-22	 	       &PrimeFullWindow     &Soft	 	& 1				\\
%  &&&&& \\
  \hline
% % % \hline
 \end{tabular}
 \label{mode}\\
  $^{a}$\citet{2015ponti_Fe}, $^{b}$\citet{2018Ponti}, $^{c}$\citet{2017Gayathri}
 \end{table*}
% +++++++++++++++++++++++++++++
%
% ***************
% 
% ^^^^^^^^^^^^^^^^^^^^
% --------------
%
% ^^^^^^^^^^^^^^^^^^^^
%
% ^^^^^^^^^^^^^^^^^^^^^^
 % ^^^^^^^^^^^^^^^^^^^^^^
 %  ============================
% % % \begin{figure*}
% % % \centering
% % % \includegraphics[scale=0.4, angle=-90]{ecp-to-OOE_flux_ratio_NH_6.ps}
% % % \caption{Line of sight column density during eclipse (top panel) and OOE  (bottom panel) phases are plotted with 
% % % X-ray reprocessing effeciency (eclipse to OOE flux ratio)}
% % % \label{fluxratioNH}
% % % \end{figure*}
%  ****************************************
 \begin{table*}
 \scriptsize
\label{table}
 \caption{The exposure times and best fit models of the eclipse and OOE spectra of four eclipsing LMXBs.
  (T$_{eff}$: Effective exposure (s), MXB: MXB 1659$-$298, EXO: EXO 0748$-$676, AXJ: AX J1745.6$-$2901, XTE: XTE J1710$-$281, E: Eclipse, OOE: Out-of-eclipse, \textit{pow}: powerlaw, \textit{ga}: Gaussian)}
 \centering
\begin{tabular}{l l l l r r}
\hline
   Source 	& State  & OB ID      & Phase  & T$_{eff}$	   & Best fit model \\
            &        &            &        &  	   	   &  \\
\hline
 MXB        & soft   & 0008620601 & OOE    & 1291  & \textit{Tbabs}$\times$\textit{xscat}$\times$\textit{gabs}(\textit{pow} + \textit{bbodyrad}) \\
             &        &            &        &  	   	   &  \\
 \hline
 MXB        & soft   &            & E      & 98    & \textit{Tbabs}$\times$\textit{xscat}$\times$\textit{pow} \\
             &        &            &        &  	   	   &  \\
\hline
 MXB        & soft   & 0008620701 & OOE    & 11747 & \textit{Tbabs}$\times$\textit{xscat}$\times$\textit{gabs$_{1}$}$\times$\textit{gabs$_{2}$}(\textit{pow} + \textit{bbodyrad}) \\
             &        &            &        &  	   	   &  \\
 \hline
 MXB        & soft   &            &  E     & 1058  & \textit{Tbabs}$\times$\textit{xscat}$\times$\textit{pow} \\
             &        &            &        &  	   	   &  \\
 \hline
 EXO        & hard   & 0160760101 & OOE    & 16378 & $\textit{Tbabs}$$\times$\textit{xscat}$\times$\textit{zxipcf}$\times$$\textit{gabs}$(\textit{pow} + \textit{diskbb} + \textit{bbodyrad}) + \textit{Tbabs}$\times$\textit{xscat}$\times$\textit{ga} \\
             &        &            &        &  	   	   &  \\
 \hline
 EXO        & hard   &            & E      & 2361  & $\textit{Tbabs}$$\times$\textit{xscat}(\textit{pow} + \textit{bbodyrad} + \textit{ga$_{1}$} + \textit{ga$_{2}$} + \textit{ga$_{3}$})\\
             &        &            &        &  	   	   &  \\
 \hline
 EXO        & hard   & 0160760201 & OOE    & 16665 & $\textit{Tbabs}$$\times$\textit{xscat}$\times$\textit{zxipcf}(\textit{pow} + \textit{diskbb}) + \textit{Tbabs}$\times$\textit{xscat}$\times$\textit{ga} \\
             &        &            &        &  	   	   &  \\
 \hline
 EXO        & hard   &            & E      & 2340  &  $\textit{Tbabs}$$\times$\textit{xscat}(\textit{pow} + \textit{bbodyrad} + \textit{ga$_{1}$} + \textit{ga$_{2}$} + \textit{ga$_{3}$} + \textit{ga$_{4}$} + \textit{ga$_{5}$})\\
             &        &            &        &  	   	   &  \\
 \hline
 EXO        & hard   & 0160760301 & OOE    & 26747 & $\textit{Tbabs}$$\times$\textit{xscat}$\times$\textit{zxipcf}(\textit{pow} + \textit{diskbb}) + \textit{Tbabs}$\times$\textit{xscat}$\times$\textit{ga} \\
             &        &            &        &  	   	   &  \\
 \hline
 EXO        & hard   &            & E      & 2662  & $\textit{Tbabs}$$\times$\textit{xscat}(\textit{pow} + \textit{bbodyrad} + \textit{ga} ) \\
             &        &            &        &  	   	   &  \\
 \hline
 EXO        & hard   & 0160760401 & OOE    & 14883 & $\textit{Tbabs}$$\times$\textit{xscat}$\times$\textit{zxipcf}$\times$\textit{gabs}(\textit{pow} + \textit{diskbb})\\
             &        &            &        &  	   	   &  \\
 \hline
 EXO        & hard   &            & E      & 1156  &  $\textit{Tbabs}$$\times$\textit{xscat}(\textit{pow} + \textit{bbodyrad} + \textit{ga}) \\
             &        &            &        &  	   	   &  \\
 \hline
 EXO        & hard   & 0160760601 & OOE    & 9094  & $\textit{Tbabs}$$\times$\textit{xscat}$\times$\textit{zxipcf}$\times$\textit{gabs}(\textit{pow} + \textit{diskbb}) + \textit{Tbabs}$\times$\textit{xscat}$\times$\textit{ga} \\
             &        &            &        &  	   	   &  \\
 \hline
 EXO        & hard   &            & E      & 1373  &  $\textit{Tbabs}$$\times$\textit{xscat}(\textit{pow} + \textit{bbodyrad} + \textit{ga})\\
             &        &            &        &  	   	   &  \\
 \hline
 EXO        & hard   & 0160760801 & OOE    & 8900  &  $\textit{Tbabs}$$\times$\textit{xscat}$\times$\textit{zxipcf}$\times$\textit{gabs}(\textit{pow} + \textit{diskbb}) + \textit{Tbabs}$\times$\textit{xscat}$\times$\textit{ga} \\
             &        &            &        &  	   	   &  \\
 \hline
 EXO        & hard   &            & E      & 1338  & $\textit{Tbabs}$$\times$\textit{xscat}(\textit{pow} + \textit{bbodyrad} + \textit{ga$_{1}$} + \textit{ga$_{2}$} + \textit{ga$_{3}$} + \textit{ga$_{4}$})\\
             &        &            &        &  	   	   &  \\
 \hline
 EXO        & hard   & 0160761301 & OOE    & 15954 & $\textit{Tbabs}$$\times$\textit{xscat}$\times$\textit{zxipcf}(\textit{pow} + \textit{diskbb})\\
             &        &            &        &  	   	   &  \\
 \hline
 EXO        & hard   &            & E      & 2327  & $\textit{Tbabs}$$\times$\textit{xscat}(\textit{pow} + \textit{bbodyrad} + \textit{ga$_{1}$} + \textit{ga$_{2}$} + \textit{ga$_{3}$})\\
             &        &            &        &  	   	   &  \\
 \hline
 AXJ        & soft   & 0402430301 & OOE    & 47261 & \textit{Tbabs}$\times$\textit{xscat}$\times$\textit{gabs$_{1}$}$\times$\textit{gabs$_{2}$}$\times$\textit{pow}\\
             &        &            &        &  	   	   &  \\
 \hline
 AXJ        & soft   &            & E      & 3292  & \textit{Tbabs}$\times$\textit{xscat}$\times$\textit{pow} \\
             &        &            &        &  	   	   &  \\
 \hline
 AXJ        & soft   & 0402430401 & OOE    & 43498 & \textit{Tbabs}$\times$\textit{xscat}$\times$\textit{gabs$_{1}$}$\times$\textit{gabs$_{2}$}(\textit{pow} + \textit{bbodyrad}) \\
             &        &            &        &  	   	   &  \\
 \hline
 AXJ        & soft   &            & E      & 3244  & \textit{Tbabs}$\times$\textit{xscat}$\times$\textit{pow} \\
             &        &            &        &  	   	   &  \\
 \hline
 AXJ        & soft   & 0724210201 & OOE    & 29217 & \textit{Tbabs}$\times$\textit{xscat}$\times$\textit{gabs$_{1}$}$\times$\textit{gabs$_{2}$}(\textit{pow} + \textit{bbodyrad}) \\
             &        &            &        &  	   	   &  \\
 \hline
 AXJ        & soft   &            & E      & 2109  & \textit{Tbabs}$\times$\textit{xscat}$\times$\textit{pow} \\
             &        &            &        &  	   	   &  \\
 \hline
 AXJ        & soft   & 0762250301 & OOE    & 83655 & \textit{Tbabs}$\times$\textit{xscat}$\times$\textit{gabs$_{1}$}$\times$\textit{gabs$_{2}$}(\textit{pow} + \textit{bbodyrad} + \textit{ga})  \\
             &        &            &        &  	   	   &  \\
 \hline
 AXJ        & soft   &            & E      & 4119  & \textit{Tbabs}$\times$\textit{xscat}$\times$\textit{pow} \\
             &        &            &        &  	   	   &  \\
 \hline
 AXJ        & hard   & 0723410301 & OOE    & 19455 & \textit{Tbabs}$\times$\textit{xscat}$\times$\textit{pow}  \\
             &        &            &        &  	   	   &  \\
 \hline
 AXJ        & hard   &            & E      & 1578  & \textit{Tbabs}$\times$\textit{xscat}$\times$\textit{pow}  \\
             &        &            &        &  	   	   &  \\
 \hline
 AXJ        & hard   & 0723410401 & OOE    & 31239 & \textit{Tbabs}$\times$\textit{xscat}$\times$\textit{pow}  \\
             &        &            &        &  	   	   &  \\
 \hline
 AXJ        & hard   &            & E      & 2208  & \textit{Tbabs}$\times$\textit{xscat}$\times$\textit{pow} \\
             &        &            &        &  	   	   &  \\
 \hline
 AXJ        & hard   & 0723410501 & OOE    & 18724 & \textit{Tbabs}$\times$\textit{xscat}$\times$\textit{pow} \\
             &        &            &        &  	   	   &  \\
 \hline
 AXJ        & hard   &            & E      & 2218  & \textit{Tbabs}$\times$\textit{xscat}$\times$\textit{pow} \\
             &        &            &        &  	   	   &  \\
 \hline
 XTE        & soft   & 0206990401 & OOE    & 5465  & \textit{Tbabs}$\times$\textit{xscat}$\times$\textit{pow} \\
             &        &            &        &  	   	   &  \\
 \hline
 XTE        & soft   &            & E      & 367   & poor statistics\textcolor{magenta}{$^{\clubsuit}$} \\
             &        &            &        &  	   	   &  \\ 
 \hline 
\end{tabular}
\label{models}\\
 \textcolor{magenta}{$\clubsuit$}: fitted with the model \textit{Tbabs}$\times$\textit{xscat}$\times$\textit{pow} only for the purpose of estimation of flux.
 \end{table*}
% *********************************************************************
\begin{table*}
 \scriptsize
 \label{table}
 \captionsetup{width=17cm}
 \caption{The best fit parameters of the eclipse and OOE spectra of the four eclipsing LMXBs, where MXB: MXB 1659$-$298, EXO: EXO 0748$-$676, AXJ: AX J1745.6$-$2901, XTE: XTE J1710$-$281, E: Eclipse, OOE: Out-of-eclipse,  N$_{\rm H}$: Line of sight equivalent hydrogen column density,  $\Gamma$: Powerlaw photon index, T$_{\rm BB}$: Blackody temperature (associated with component \textit{bbodyrad}),  T$_{\rm in}$: Inner accretion disk temperature (associated with   component \textit{diskbb}), N$_{\rm bbrd}$: Blackody normalization associated with model component \textit{bbodyrad} (($\frac{\rm R_{\rm km}}{\rm D_{10} \rm kpc})^{2}$), N$_{\rm disk}$: Blackody normalization associated with model component \textit{diskbb} (($\frac{\rm R_{\rm in}}{\rm D_{10}})^{2}$$\rm cos\theta$), DOF: Degrees of freedom.  Flux ratio gives the ratio of (3-10 keV) flux in the OOE phase with that in eclipse phase. R$_{\rm km}$, R$_{\rm in}$ and $\theta$ are the radius of the blackbody emitting region, apparent inner accretion disk radius and  angle of the normal to the disk plane with the line of sight.
 The errors are given with 90$\%$ confidence limit. frz: Frozen }
 \centering
 \begin{tabular}{lll l ll ll lc r}
\hline
Source 	&State & OB ID & Phase & N$_{\rm H}$  		   & $\Gamma$     &  T$_{\rm BB}$ or T$_{\rm in}$ 	 &  N$_{\rm bbrd}$	or N$_{\rm disk}$         & $\chi^{2}$/DOF & Flux             & Flux\\
        &	   &	   &       & (10$^{22}$ cm$^{-2}$) &  		      &  (keV)	         &               &                & (3-10 keV)       & ratio \\
        &      &       &       &                       &              &                  &                          &                & (10$^{-11}$ ergs &  \\
        &      &       &       &                       &              &                  &                          &                & cm$^{-2}$ sec$^{-1}$)             &  \\
\hline
 MXB  & soft & 0008620601 & OOE & 0.28$^{+0.01}_{-0.01}$ & 1.75$^{+0.04}_{-0.03}$ & 1.10$^{+0.06}_{-0.05}$ [T$_{\rm BB}$] & 12.06$^{+2.01}_{-2.04}$ & 142.21/162 & 55.90 & 21.09\\
\hline
      &      &            & E   & 0.25$^{+0.08}_{-0.07}$ & 1.52$^{+0.21}_{-0.20}$ &  $-$                   &   $-$                   & 21.75/25   & 2.65  & \\
 && && && && && \\
\hline
 MXB  & soft & 0008620701 & OOE & 0.35 (frz)          & 2.07$^{+0.18}_{-0.13}$ & 1.75$^{+0.12}_{-0.15}$ [T$_{\rm BB}$] & 2.77$^{+0.54}_{-0.36}$  & 96.85/105  & 61.14 & 23.88 \\
\hline
      &      &            & E   & 0.26$^{+0.02}_{-0.02}$ & 1.57$^{+0.05}_{-0.05}$ &   $-$               &   $-$                   & 94.47/105  & 2.48   & \\
 && && && && && \\
\hline 
 EXO & hard  & 0160760101 & OOE & 0.11 (frz)          & 2.24$^{+0.11}_{-0.12}$ & 0.09$^{+0.00}_{-0.01}$ [T$_{\rm in}$]& 180055 (frz)  & 195.87/166 & 17.84  & 54.06 \\
\hline
     &       &             & E   & 0.04$^{+0.01}_{-0.01}$ & 1.04$^{+0.17}_{-0.18}$ & 0.21$^{+0.02}_{-0.03}$ [T$_{\rm BB}$] & 24  (frz)            & 67.14/70   &  0.33  & \\
 && && && && && \\
\hline
 EXO  & hard & 0160760201 & OOE & 0.11 (frz)          & 1.31$^{+0.01}_{-0.01}$ & 0.14$^{+0.01}_{-0.01}$ [T$_{\rm in}$] & 18971$^{+4210}_{-3164}$ & 234.28/167 &  14.77 & 49.23 \\
\hline
      &      &            & E   & 0.03 (frz)          & 1.14$^{+0.15}_{-0.15}$ & 0.19$^{+0.01}_{-0.01}$ [T$_{\rm BB}$] & 68  (frz)            & 56.51/55   & 0.30   & \\
 && && && && && \\
\hline
  EXO & hard & 0160760301 & OOE & 0.11 (frz)          & 1.34$^{+0.01}_{-0.01}$ & 0.15$^{+0.00}_{-0.00}$ [T$_{\rm in}$] & 18073$^{+2777}_{-2271}$ & 142.81/168 & 15.82  & 46.53 \\
\hline
      &      &            & E   & 0.03 (frz)          & 1.06$^{+0.14}_{-0.14}$ & 0.17$^{+0.01}_{-0.01}$ [T$_{\rm BB}$] &  155$^{+42}_{-32}$     & 97.58/84  & 0.34   & \\
 && && && && && \\
\hline
  EXO & hard & 0160760401 & OOE & 0.11 (frz)          & 1.39$^{+0.02}_{-0.02}$ & 0.20$^{+0.01}_{-0.01}$ [T$_{\rm in}$] & 13894$^{+9526}_{-6213}$ & 166.44/150 & 13.84  & 49.43 \\
\hline
      &      &            & E   & 0.03 (frz)          & 0.87$^{+0.18}_{-0.18}$ & 0.13$^{+0.00}_{-0.00}$ [T$_{\rm BB}$] & 389 (frz)               & 46.29/34   & 0.28   & \\
 && && && && && \\
\hline
  EXO & hard & 0160760601 & OOE & 0.11 (frz)          & 1.32$^{+0.02}_{-0.02}$  & 0.15$^{+0.01}_{-0.01}$ [T$_{\rm in}$] & 14310$^{+2666}_{-2129}$ & 219.38/166 & 15.64  & 53.93 \\
\hline
      &      &            & E   & 0.03 (frz)          & 0.92$^{+0.17}_{-0.18}$  & 0.15$^{+0.01}_{-0.01}$ [T$_{\rm BB}$] & 234$^{+91}_{-62}$       & 37.90/40   & 0.29   & \\
 && && && && && \\
\hline
  EXO & hard & 0160760801 & OOE & 0.11 (frz)          & 1.27$^{+0.01}_{-0.01}$  & 0.14$^{+0.01}_{-0.01}$ [T$_{\rm in}$] & 11891$^{+3686}_{-2473}$ & 193.93/165 & 14.41  & 43.67 \\
\hline
      &      &            & E   & 0.03 (frz)          & 1.16$^{+0.14}_{-0.14}$  & 0.09$^{+0.00}_{-0.00}$ [T$_{\rm BB}$] & 1268 (frz)           & 29.36/42   & 0.33   & \\
 && && && && && \\
\hline
  EXO & hard & 0160761301 & OOE & 0.11 (frz)          & 1.32$^{+0.02}_{-0.02}$  & 0.25$^{+0.01}_{-0.02}$ [T$_{\rm in}$] & 1041$^{+216}_{-181}$    & 183.23/153 & 14.59  & 42.91 \\
\hline
      &      &            & E   & 0.03 (frz)          & 1.14$^{+0.10}_{-0.10}$  & 0.13$^{+0.00}_{-0.00}$ [T$_{\rm BB}$] & 318 (frz)            & 51.95/66   & 0.34   & \\
 && && && && && \\
\hline
 AXJ  & soft & 0402430301 & OOE & 18.64$^{+0.19}_{-0.19}$ & 2.36$^{+0.02}_{-0.02}$ & $-$                    & $-$                     & 145.68/105 & 13.28 & 6.48\\
\hline
      &      &            & E   & 17.14$^{+1.05}_{-0.99}$ & 3.36$^{+0.17}_{-0.16}$ & $-$                    & $-$                     & 100.59/86  & 2.05 & \\
 && && && && && \\
\hline
 AXJ  & soft & 0402430401 & OOE & 16.86$^{+0.81}_{-0.81}$ & 1.65$^{+0.16}_{-0.21}$        & 1.17$^{+0.04}_{-0.04}$ [T$_{\rm BB}$] & 8.50 (frz)  & 157.33/104 & 14.61 & 8.12 \\
\hline
      &      &            & E   & 19.23$^{+1.18}_{-1.11}$ & 3.83$^{+0.20}_{-0.19}$ & $-$                    & $-$                     & 85.80/80   &  1.80 & \\
 && && && && && \\
 \hline
  AXJ & soft & 0724210201 & OOE & 20.58$^{+0.43}_{-0.46}$ & 2.25 (frz)          & 1.53$^{+0.07}_{-0.10}$ [T$_{\rm BB}$] & 1.16$^{+0.31}_{-0.32}$  & 121.33/104 & 15.03 & 6.37\\
\hline
      &      &            & E   & 17.83$^{+1.27}_{-1.19}$ & 3.03$^{+0.19}_{-0.19}$ & $-$                    & $-$                     & 91.19/77   & 2.36 & \\
 && && && && && \\
\hline  
  AXJ & soft & 0762250301 & OOE & 16.35$^{+0.88}_{-1.07}$ & 2.09 (frz)          & 1.77$^{+0.02}_{-0.02}$ [T$_{\rm BB}$] & 2.98$^{+0.45}_{-0.44}$  & 132.08/105 &  19.57 & 8.98\\
 \hline
      &      &            & E   & 18.03$^{+2.65}_{-2.37}$ & 3.33$^{+0.48}_{-0.44}$ &  $-$                   & $-$                     & 42.53/50   &  2.18& \\
 && && && && && \\
\hline
  AXJ & hard & 0723410301 & OOE & 19.53$^{+0.88}_{-4.14}$ & 1.82$^{+0.08}_{-0.08}$ &   $-$                  & $-$                     & 103.05/101 & 2.13 & 9.68\\
\hline
      &      &            & E   & 9.39 (frz)           & 2.05$^{+0.56}_{-0.55}$ &   $-$                  & $-$                     & 4.87/8 & 0.22 & \\
 && && && && && \\
 \hline
  AXJ & hard & 0723410401 & OOE & 17.15$^{+0.84}_{-0.83}$ & 1.65$^{+0.08}_{-0.08}$ &   $-$                  & $-$                     & 91.22/102 & 1.34 & 10.31\\
 \hline 
       &     &            & E   &  17.15 (frz)         & 2.65$^{+0.10}_{-0.08}$ &   $-$                  & $-$                     &  3.63/5 &  0.13 & \\
 && && && && && \\
\hline
  AXJ & hard & 0723410501 & OOE & 19.50$^{+1.03}_{-1.01}$ & 1.89$^{+0.10}_{-0.10}$ &  $-$                   & $-$                     & 74.65/99 & 1.43 & 5.72\\
\hline
      &      &            & E   & 19.50 (frz)          & 2.39$^{+0.52}_{-0.54}$ &   $-$                  & $-$                     & 10.74/10 & 0.25 & \\
 && && && && && \\
 \hline
 XTE  & soft & 0206990401 & OOE & 0.27$^{+0.01}_{-0.01}$  & 1.92$^{+0.03}_{-0.02}$ &   $-$                  & $-$                     & 197.74/150 & 2.58 & 129 \\
\hline
      &      &            & E   &   $-$                   &  $-$                   &   $-$                  &  $-$                    &   $-$             & 0.02 & \\
 && && && && && \\
\hline
 \end{tabular}
 \label{phenomeno}\\
  \end{table*}
%
% =======================
\begin{table*}
\scriptsize
\label{table}
% \centering
% \captionsetup{width=16.5cm}
\captionsetup{width=18cm}
\caption{The emission and absorption line parameters of the eclipse and OOE spectra of the three eclipsing LMXBs, 
 MXB: MXB 1659$-$298, EXO: EXO 0748$-$676, AXJ: AX J1745.6$-$2901, XTE: XTE J1710$-$281, E: Eclipse, OOE: Out-of-eclipse, E$_{em}$:  Emission line energy (keV), w$_{em}$: Emission line width (keV),  F$_{em}$: Emission line flux (photons cm$^{-2}$ sec$^{-1}$),  EqW: Equivalent width of emission line (eV), E$_{ab}$:  Absorption line energy (keV), w$_{ab}$: Absorption line width (keV), S$_{ab}$: Absorption line strength. Errors in line energies and widths are given with 90$\%$ confidence limit, whereas that on emission line flux, equivalent width of the emission line and absorption line strength are given at 1$\sigma$ confidence limit.}
 \centering
\begin{tabular}{l l l l l l l l l l l}
\hline
\hline
 Source & State & OB ID      & Phase & E$_{em}$ 	  & w$_{em}$  	 	   & F$_{em}$		              &    EqW  	    &  E$_{ab}$              & w$_{ab}$      & S$_{ab}$  \\
  &  & &  & &  & &  & &  & \\
 \hline
 \hline
 MXB    & soft  & 0008620601 & OOE   &   $- $         &    $- $            &     $- $                     &     $- $        & 6.89$^{+0.11}_{-0.10}$ & 0.11 (frz) & 0.04$^{+0.01}_{-0.01}$\\
 \hline
 MXB    & soft  & 0008620701 & OOE   &   $- $         &   $- $             &     $- $                     &     $- $        & 6.71$^{+0.03}_{-0.02}$ & 0.0064  (frz) & 0.075$^{+0.05}_{-0.03}$\\
 \hline
        &      &             & OOE   &   $- $         &   $- $             &     $- $                     &     $- $        & 6.99$^{+0.02}_{-0.02}$ & 0.0077   (frz) & 0.073 (frz) \\
 \hline
 EXO    & hard  & 0160760101 & OOE   & 0.525 (frz)         & 0.14$^{+0.01}_{-0.01}$ & 0.01$^{+0.001}_{-0.001}$ & 58$^{+6}_{-6}$    & 6.75$^{+0.13}_{-0.13}$ & 0.19 (frz)  & 0.02$^{+0.01}_{-0.01}$\\
 \hline
        &       &            & E     & 1.40$^{+0.03}_{-0.03}$ & 0.000096 (frz)   & 0.000017$^{+0.000006}_{-0.000006}$ & 58$^{+20}_{-20}$  &     $- $  &      $- $  &   $- $ \\
 \hline
        &       &            & E     & 0.89$^{+0.01}_{-0.01}$ & 0.03 (frz)          & 0.000093$^{+0.00001}_{-0.00001}$   & 138$^{+15}_{-15}$ &     $- $ &    $- $  &     $- $\\
 \hline
        &       &            & E     & 0.51$^{+0.02}_{-0.02}$ & 0.11$^{+0.02}_{-0.02}$ & 0.00095 (frz)                   & 708  &    $- $  &     $- $  &     $- $ \\
 \hline
 EXO    & hard  & 0160760201 & OOE   & 0.60$^{+0.01}_{-0.01}$ & 0.07$^{+0.01}_{-0.01}$ & 0.004$^{+0.001}_{-0.001}$ & 42$^{+10}_{-10}$ &     $- $ &     $- $  &    $- $  \\
 \hline
       &        &            & E    & 1.39$^{+0.03}_{-0.03}$ & 0.000015 (frz)      & 0.000019$^{+0.000006}_{-0.000006}$ & 61$^{+19}_{-19}$ &    $- $ &    $- $  &     $- $\\ 
\hline
       &        &            & E    & 1.80$^{+0.04}_{-0.03}$ & 0.0000032  (frz)    & 0.000013$^{+0.000005}_{-0.000005}$ & 72$^{+28}_{-28}$  &    $- $ &    $- $  &     $- $ \\ 
 \hline
       &        &            & E    & 6.44$^{+0.06}_{-0.08}$ & 0.000062  (frz)     & 0.000013$^{+0.000005}_{-0.000005}$ & 335$^{+129}_{-129}$ &   $- $ &    $- $ &    $- $ \\ 
 \hline
       &         &           & E    & 0.89$^{+0.02}_{-0.01}$ & 0.03 (frz)          & 0.000086$^{+0.00001}_{-0.00001}$   & 101$^{+12}_{-12}$ &   $- $  &    $- $  &   $- $  \\ 
 \hline
       &         &           & E    & 0.54$^{+0.01}_{-0.02}$ & 0.06$^{+0.02}_{-0.02}$ & 0.00037$^{+0.00005}_{-0.00005}$    & 201$^{+27}_{-27}$ &   $- $ &   $- $  &   $- $ \\ 
 \hline
 EXO    & hard  & 0160760301 &  OOE & 0.62$^{+0.01}_{-0.01}$ & 0.04$^{+0.02}_{-0.02}$ & 0.002$^{+0.0003}_{-0.0003}$          & 14$^{+2}_{-2}$  &   $- $ &    $- $ &    $- $ \\
  \hline
        &       &             &  E & 0.55$^{+0.02}_{-0.02}$  & 0.04$^{+0.03}_{-0.04}$ & 0.00018 (frz)          & 69  &   $- $ &    $- $ &    $- $ \\
 \hline
  EXO    & hard & 0160760401  &  OOE &   $- $                &   $- $                  &   $- $                             &   $- $         & 6.85$^{+0.12}_{-0.13}$ &  0.14 (frz) & 0.013$^{+0.004}_{-0.004}$ \\
 \hline
         &      &             &   E  &  0.90$^{+0.06}_{-0.03}$ & 0.000056  (frz)       & 0.000037$^{+0.000014}_{-0.000013}$ & 64$^{+24}_{-22}$ &  $- $ &  $- $ &  $- $ \\
 \hline
  EXO    & hard & 0160760601  &  OOE & 0.59$^{+0.01}_{-0.01}$ & 0.05$^{+0.01}_{-0.01}$ & 0.002$^{+0.000}_{-0.000}$ & 22$^{+0}_{-0}$ & 6.9 (frz) &  0.18$^{+0.13}_{0.10-}$ & 0.02$^{+0.00}_{-0.00}$ \\
 \hline
         &      &             &  E   & 1.89 (frz)          & 0.11$^{+0.09}_{-0.05}$ & 0.00003 (frz)          & 236 &    $- $ &   $- $   &    $- $ \\
 \hline
  EXO    & hard & 0160760801  &  OOE & 0.58$^{+0.01}_{-0.02}$ & 0.07$^{+0.02}_{-0.01}$ & 0.003$^{+0.00}_{-0.00}$   & 129$^{+0}_{-0}$ & 6.81$^{+0.08}_{-0.08}$ &  0.15$^{+0.06}_{-0.09}$ & 0.02$^{+0.00}_{-0.00}$ \\
 \hline
         &      &             &  E   & 1.28$^{+0.04}_{-0.05}$         & 0.06 (frz)           & 0.000036$^{+0.00001}_{-0.00001}$        & 129$^{+36}_{-36}$    &    $- $ &    $- $  &   $- $ \\
 \hline
         &      &             &  E   & 6.40 (frz)          &  0.19 (frz)          & 0.00003$^{+0.000009}_{-0.000009}$         & 655$^{+196}_{-196}$ &    $- $ &     $- $ &    $- $ \\
 \hline
         &      &             &  E   & 0.595$^{+0.02}_{-0.02}$ & 0.06$^{+0.03}_{-0.02}$ & 0.00027 (frz)         & 192 &   $- $  &    $- $  &   $- $ \\
 \hline
         &      &             &  E   & 0.89$^{+0.03}_{-0.03}$  & 0.07$^{+0.03}_{-0.02}$ & 0.00013 (frz)                  & 262 &   $- $  &    $- $ &   $- $ \\
 \hline
  EXO    & hard & 0160761301  &  E   & 1.32$^{+0.03}_{-0.04}$  & 0.00016 (frz)       & 0.000019$^{+0.000006}_{-0.000006}$ & 63$^{+21}_{-21}$  &  $- $   &   $- $ &   $- $ \\
 \hline
         &      &             &  E   & 0.55$^{+0.01}_{-0.01}$ & 0.002 (frz) & 0.00021$^{+0.000036}_{-0.000036}$ & 92$^{+16}_{-16}$ &  $- $ &   $- $ &   $- $ \\
 \hline
         &      &             &  E   & 0.90 $^{+0.02}_{-0.02}$ & 0.03 (frz) & 0.000060$^{+0.00001}_{-0.00001}$ & 87$^{+18}_{-18}$ &  $- $ &  $- $  &  $- $ \\
 \hline
  AXJ    & soft & 0402430301  & OOE  &   $- $ &    $- $ &   $- $ &   $- $  & 6.68$^{+0.03}_{-0.03}$ & 0.094 (frz) & 0.035$^{+0.004}_{-0.003}$ \\
 \hline
         &      &             & OOE  &  $- $  &   $- $  &  $- $  &   $- $  & 6.95$^{+0.02}_{-0.02}$ & 0.02 (frz) & 0.049$^{+0.005}_{-0.005}$ \\
 \hline
  AXJ    & soft & 0402430401  & OOE  &   $- $ &  $- $   &   $- $ &   $- $  & 6.66$^{+0.03}_{-0.02}$ & 0.01 (frz) & 0.03$^{+0.01}_{-0.00}$ \\
 \hline
         &      &             & OOE  &   $- $ &  $- $   &   $- $ &   $- $  & 6.96$^{+0.01}_{-0.01}$ & 0.009$^{+0.001}_{-0.001}$ & 0.15 (frz) \\
 \hline
  AXJ    & soft & 0724210201  & OOE  &   $- $ &    $- $ &   $- $ &   $- $ & 6.71$^{+0.03}_{-0.03}$ & 0.004 (frz)  & 1.03  (frz) \\
 \hline
         &      &             & OOE  &   $- $ &    $- $ &   $- $ &   $- $ & 6.96$^{+0.03}_{-0.02}$ & 0.01$^{+0.00}_{-0.00}$ & 0.32  (frz) \\
 \hline
  AXJ    & soft & 0762250301  & OOE  & 6.4 (frz) &  0.002 (frz)  & 0.000042$^{+0.00001}_{-0.00001}$ & 8$^{+2}_{-2}$ & 6.71$^{+0.04}_{-0.04}$ & 0.002 (frz) & 0.06 (frz)  \\
 \hline
         &      &             & OOE  &    $- $ &    $- $  &   $- $  &   $- $  & 6.96$^{+0.02}_{-0.01}$ & 0.004 (frz) & 5.76 (frz) \\
 \hline
 \hline
% % %  \end{tabular}
 \label{phenomeno_line}
 \\
\end{tabular}
 \end{table*}
%
%  ------------------------
% =======================
\begin{table*}
\footnotesize
\label{table}
 \caption{\textit{zxipcf} parameters for the OOE spectra of EXO 0748$-$676, where  
 N$_{\rm h}$: Line of sight equivalent hydrogen column density associated with model component \textit{zxipcf}, log($\xi$): Logarithm of ionization parameter $\xi$,  CvF: covering fraction. Errors are given with 90$\%$ confidence limit.}
 \centering
 \begin{tabular}{lc l c}
 \hline
 
 OB ID          &  N$_{\rm h}$ 	           & log($\xi$)	             & CvF     \\
                & (10$^{22}$ cm$^{-2}$)    &                         &         \\
 \hline
  0160760101    &  5.35$^{+0.32}_{-0.24}$  &  1.37$^{+0.19}_{-0.10}$ &  0.76$^{+0.03}_{-0.04}$      \\
  \hline
  0160760201    &  5.62$^{+0.72}_{-0.58}$  &  2.02$^{+0.03}_{-0.04}$ &  0.61$^{+0.04}_{-0.04}$      \\ 
  \hline  
  0160760301    &  4.18$^{+0.47}_{-0.37}$  &  2.05$^{+0.03}_{-0.03}$ &  0.55$^{+0.03}_{-0.04}$      \\
  \hline
  0160760401    &  2.14$^{+0.18}_{-0.17}$  &  0.59$^{+0.10}_{-0.11}$ &  0.92$^{+0.03}_{-0.04}$      \\
  \hline  
  0160760601    &  4.64$^{+0.35}_{-0.29}$  &  1.97$^{+0.02}_{-0.02}$ &  0.72$^{+0.03}_{-0.03}$     \\
  \hline
  0160760801    &  6.35$^{+0.56}_{-0.46}$  &  1.96$^{+0.02}_{-0.02}$ &  0.76$^{+0.03}_{-0.04}$      \\
  \hline  
  01607601301   &  3.86 (frz)              &  1.28$^{+0.17}_{-0.18}$ &  0.79$^{+0.05}_{-0.05}$      \\
  \hline
\hline
\label{EXO_OOE_zxipcf}
 \end{tabular}
 \end{table*}
% ======================= 
%  ------------------------
%
\begin{table*} 
\scriptsize
 \label{table}
% \centering
% \captionsetup{width=16.5cm}
\caption{0.3-3 keV flux obtained from the best fit eclipse spectra of the three eclipsing LMXBs including the dust scattering  model component
 \textit{xscat} (F$_{\rm low\_xs}$) and that excluding \textit{xscat} (F$_{\rm low}$). Last two columns give the reduced $\chi^{2}$ of the best fit eclipse spectra including  and excluding the model component \textit{xscat} respectively. MXB: MXB 1659$-$298, EXO: EXO 0748$-$676, AXJ: AX J1745.6$-$2901}
 \centering
\begin{tabular}{l l l c c l l}
\hline
 Source & State & OB ID       & F$_{xs}$ (0.3-3 keV)               & F (0.3-3 keV)                      & $\chi_{xs}^{2}$/DOF  & $\chi^{2}$/DOF \\
        &       &             & (10$^{-11}$ ergs cm$^{-2}$ s$^{-1}$)  & (10$^{-11}$ ergs cm$^{-2}$ s$^{-1}$) &                      &      \\          
 \hline
 \hline
   &&&&&& \\
 MXB    & soft  & 0008620601  & 1.3285                         & 1.3277                        & 21.75/25 (0.87)      & 21.76/25 (0.8704) \\
 \hline
  &&&&&& \\
 MXB    & soft  & 0008620701  & 1.3553                         & 1.3558                        & 103.24/106 (0.9740)  & 103.04/106 (0.9721) \\
 \hline
  &&&&&& \\
 EXO    & hard  & 0160760101  & 0.2288                         & 0.2286                        & 67.14/70 (0.9591)     & 67.12/70 (0.9588) \\
 \hline
  &&&&&& \\
 EXO    & hard  & 0160760201  & 0.2314                         & 0.2317                        & 56.51/55 (1.0274)     & 56.52/55 (1.0276) \\
 \hline
  &&&&&& \\
 EXO    & hard  & 0160760301  & 0.2338                         & 0.2338                        & 97.58/84 (1.1617)     & 97.46/84 (1.1602) \\
 \hline
  &&&&&& \\
 EXO    & hard  & 0160760401  & 0.1617                         & 0.1624                        & 46.29/34 (1.3615)     & 46.25/34 (1.3603) \\
 \hline
  &&&&&& \\
 EXO    & hard  & 0160760601  & 0.1883                         & 0.1884                        & 37.90/40 (0.9475)     & 37.78/40 (0.9445) \\
 \hline
  &&&&&& \\
 EXO    & hard  & 0160760801  & 0.2195                         & 0.2196                        & 29.36/42 (0.6990)     & 29.31/42 (0.6978) \\
 \hline
  &&&&&& \\
 EXO    & hard  & 0160761301  & 0.2258                         & 0.2262                        & 51.95/66 (0.7871)     & 52.28/66 (0.7921) \\
 \hline
  &&&&&& \\
 AXJ    & soft  & 0402430301  & 0.0742                         & 0.0742                        & 100.59/87 (1.1562)    & 100.58/87 (1.1561) \\
 \hline
  &&&&&& \\
 AXJ    & soft  & 0402430401  & 0.0552                         & 0.0557                        & 85.80/80 (1.0725)     & 85.37/80 (1.0671) \\
 \hline
  &&&&&& \\
 AXJ    & soft  & 0724210201  & 0.0511                         & 0.0515                        & 91.19/77 (1.1843)     & 90.45/77 (1.1747) \\
 \hline
  &&&&&& \\
 AXJ    & soft  & 0762250301  & 0.0569                         & 0.0575                        & 42.53/50 (0.8506)     & 42.37/50 (0.8474)\\
 \hline
  &&&&&& \\
 AXJ    & hard  & 0723410301  & 0.0113                         & 0.0115                        & 4.87/8 (0.6087)       & 4.87/8 (0.6087) \\
 \hline
  &&&&&& \\
 AXJ    & hard  & 0723410401  & 0.0027                         & 0.0028                        & 3.63/5 (0.726)        & 3.62/5 (0.724) \\
 \hline
  &&&&&& \\
 AXJ    & hard  & 0723410501  & 0.0027                         & 0.0028                        & 10.74/10 (1.074)     & 10.74/10 (1.074) \\
\hline 
\hline
\end{tabular}
 \label{compare_xscat_NOxscat}
 \\
 \end{table*}
%  ------------------------
% ^^^^^^^^^^^^^^^^^^^^^^^^^^^^^^^^^^^^^
\begin{figure*}
\centering
% % % \includegraphics[scale=0.25, angle=0]{three_occul_in_one_fig_arrow.eps}
\includegraphics[scale=0.25, angle=0]{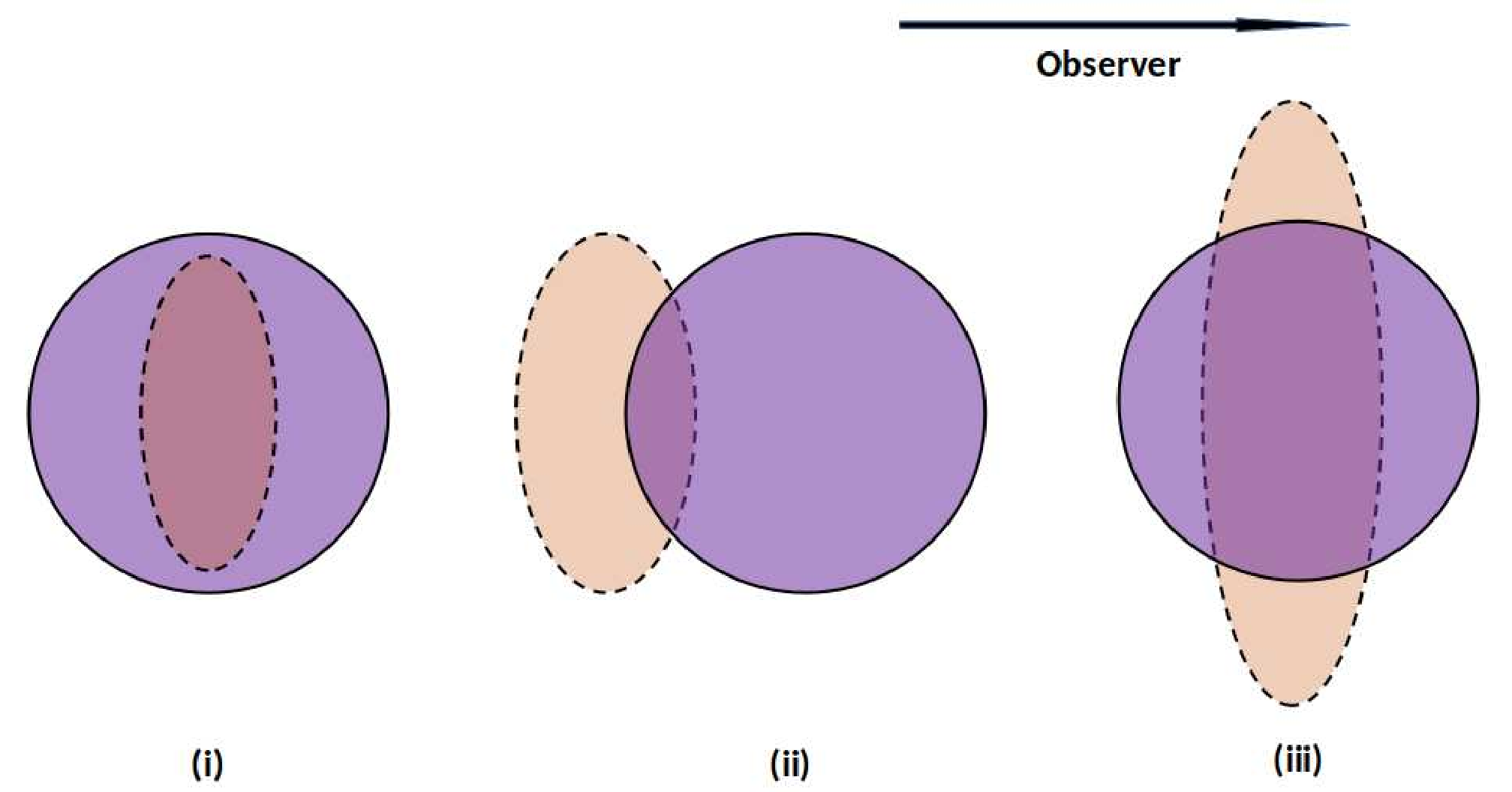}
\caption{The types of occultation during maximum eclipse phase depending upon the inclination $i$, radius of the companion star R$_{\rm C}$, radius of the accretion disk R$_{\rm out}$ and the binary separation $a$. (i) The projection of the accretion disk (ellipse) is fully covered by the projection of the companion star (circle). (ii) The projection of the accretion disk is partially covered and intersected by the projection  of the companion star at two points. (iii) The projection of the accretion disk is partially covered and  intersected by the projection  of the companion star at four points.}
\label{three_tpyes_occult}
\end{figure*}
% ===========================================
% ---------------------------------------------------
% ===========================================
\begin{figure*}
\centering
\includegraphics[scale=0.8, angle=0]{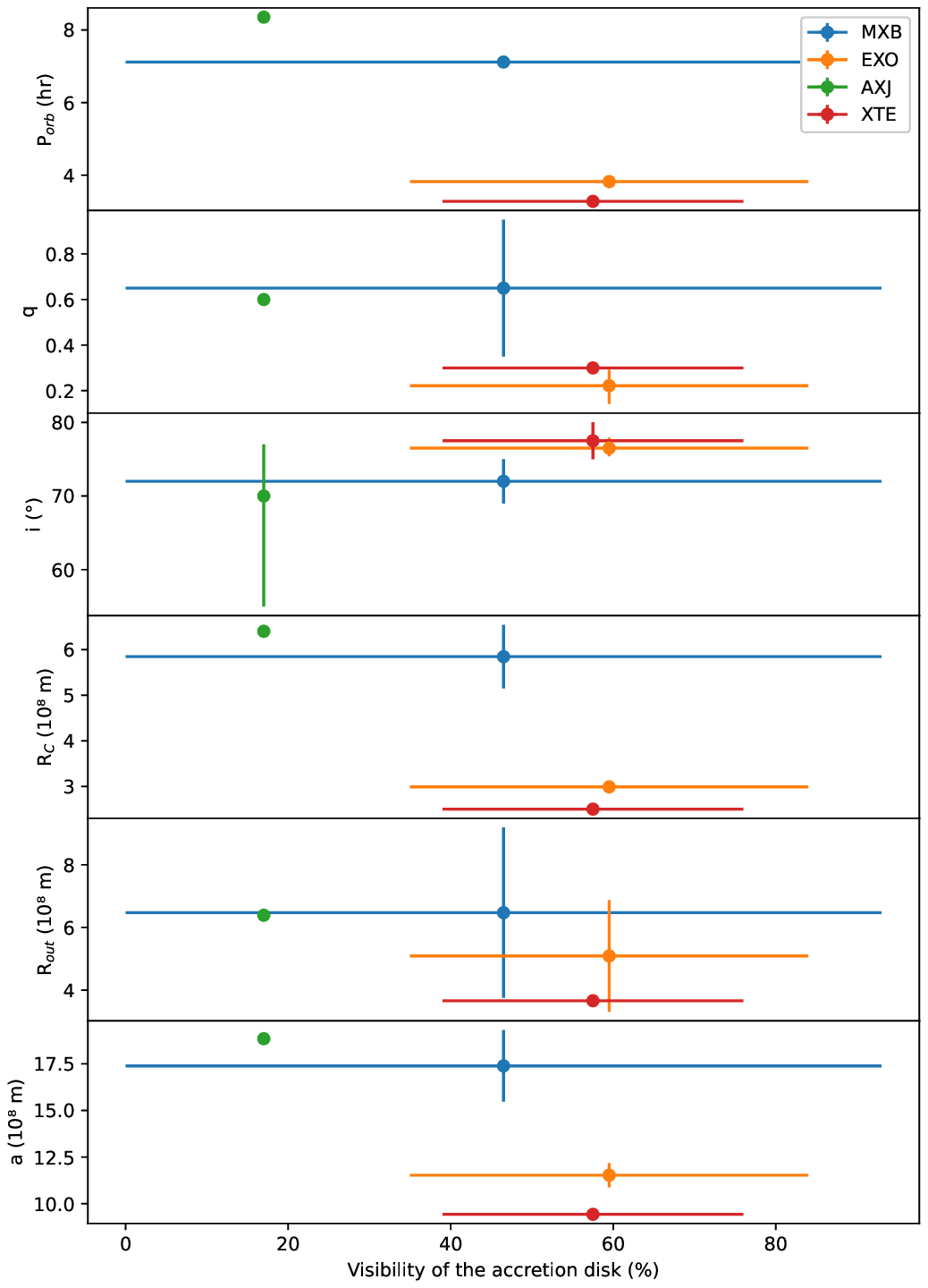}
\caption{The orbital parameters, radius of the companion star and accretion disk  are plotted against visible (during  maximum eclipse phase) portion of the  accretion disk with respect to it's total area in $\%$. P$_{\rm orb}$: orbital period of the binary, q: ratio of companion star's mass with neutron star's mass, i: inclination of the system, R$_{\rm C}$: radius of the companion star,  R$_{\rm out}$: radius of the  accretion disk, a: binary separation.}
\label{June15_Vis_six_panel}
\end{figure*}
% ===========================================
% ---------------------------------------------------
% ===========================================
% ---------------------------------------------------
% ===========================================
\begin{figure*}
\centering
\includegraphics[scale=0.8, angle=0]{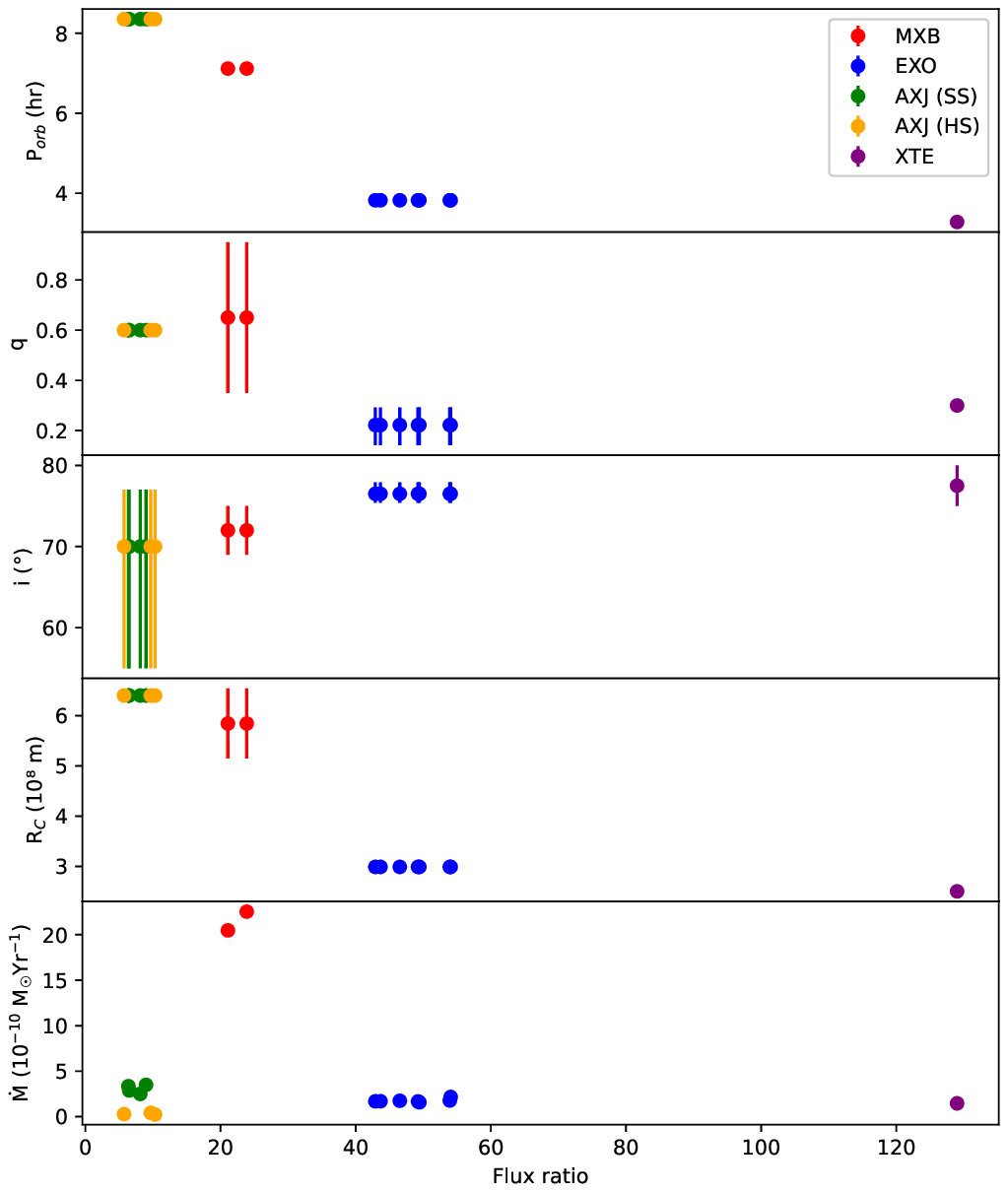}
\caption{The orbital parameters, radius of the companion and mass accretin rate are plotted against
OOE to eclipse flux ratio. P$_{\rm orb}$: orbital period of the binary, q: ratio of companion star's mass with neutron star's mass, i: inclination of the system, R$_{\rm C}$: radius of the companion star,  $\dot{\rm M}$: Mass accreiton rate.}
\label{June15_fratio_five_panel}
\end{figure*}
% ===========================================
% ---------------------------------------------------
% ===========================================
\begin{figure*}
\centering
\includegraphics[scale=0.8, angle=0]{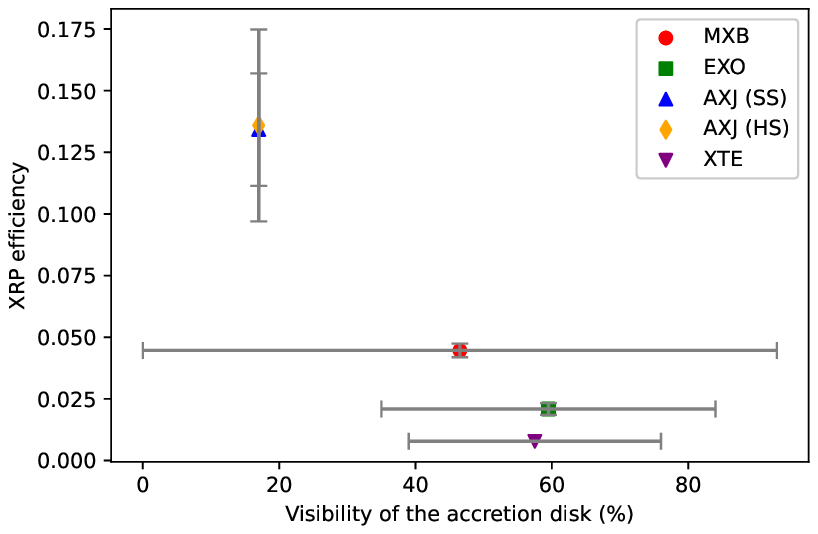}
\caption{The X-ray reprocessing efficiency (reciprocal of the ratio of OOE flux to eclipse flux) is plotted against visible (during total eclipse) portion of the  accretion disk with respect to it's total area in $\%$. }
\label{June15_Vis_XRP}
\end{figure*}
% ===========================================
% ---------------------------------------------------
% ===========================================
\section{Discussion}
During eclipses in LMXB systems the direct emission is completely blocked by the companion and the 
detected X-rays are the reprocessed X-rays from the surrounding medium. In LMXBs the main reprocessing
agent is the accretion disk and any structures if present on the disk. 
Corona above the accretion disk and outflows from disk also act as reprocessing agents in the systems where
these are present. In some cases 
slow and dense wind from the red giant companion (\citealt{2007Masetti}, \citealt{2010Nespoli})
acts as a reprocessing agent. 
During eclipses the continuum is suppressed, because the direct emission from the neutron star is blocked
and the emission lines
which are often produced in the surrounding medium are expected to be suppressed by a lesser extent \citep{Aftab2019}.
The eclipse spectrum is therefore expected to show larger equivalent width and better detectability of the 
emission lines compared to the OOE spectrum, if the emitting plasma is present in the medium surrounding the central source.
Absorption lines in  LMXBs are produced during the passage of the photons through the accretion disk 
atmosphere (\citealt{Roza_etal_2014}, \citealt{Trigo_Boirin_16}).
So their detectability is better during OOE phase.
  However depending upon the extent of the accretion disk; its orientation with respect to the line of sight; structures
  in the disk; distribution of the material 
around the compact object and its density, 
the eclipse spectrum in comparison to the OOE spectrum can be different in different sources and also
in the same source at different epochs.
% ---------------
\par
% ----------
  \par
  To understand the X-ray reprocessing properties of LMXB systems we have analyzed 17 observations of 4 eclipsing
  LMXB sources with \textit{XMM-Newton} EPIC pn. 
  These LMXBs have orbital period in the range of 3.281 hour to 8.351 hour, so far known mass of the companion
is in the range of 0.2M$_{\odot}$ to 1.0M$_{\odot}$ 
(EXO 0748$-$676: \citealt{2009Hynes}, AX J1745.6$-$2901: \citealt{1996Maeda} and MXB 1659$-$298: \citealt{2018Ponti}).
  We have compared eclipse spectra of these sources and
  also compared eclipse and OOE spectra of each observation. 
  We have obtained flux during eclipse and OOE phase in 
  each observation from the best fit models. We have also estimated the flux by fitting each spectrum with a simple powerlaw modified with
  photoelectric absorption and found the same value of flux that obtained with best fit model. This signifies that the quoted flux
  are independent of the choice of models. We have measured the X-ray flux of theses sources during eclipses and have made a comparison with the OOE fluxes. This is to be seen in the context of each binary separately, specially in context with their orbital parameters, for example, size of the binary,
relative size of the accretion disk with companion star’s, for which the binary parameters of each source is shown Table \ref{intro} and \ref{three_distances}.
\par
 Two observations of MXB 1659$-$298 observed in soft states show nearly same fluxes 
 in the energy range of (0.3-10) keV in the eclipse phases.
 Flux ratio of the OOE to eclipse phases are also quite similar, which signifies that
 X-ray reprocessing properties of the system in the two soft states of the source, observed 
 within a gap of $\sim$1 year are consistent.
 N$_{\rm H}$ obtained from the eclipse and OOE spectra of the two observations do not vary
much, this signifies nearly consistent environment surrounding the neutron star.
  Black body temperature found in the two observations in OOE phase are high ($>$1 keV). Considering the
 distance of the system to be $10-16$ kpc \citep{2006Trigo} and from the blackbody 
 normalization (($\frac{\rm R_{\rm km}}{\rm D_{10} \rm kpc})^{2}$: $2.41-14.07$, Table \ref{phenomeno}),
 the size (radius) of the emitting region is estimated as
 $\sim$$1.6-6$ km.
In the composite eclipse (OOE) spectra we see both the eclipse (OOE)
spectra almost overlap (Fig \ref{MXB_comp}), that indicates nearly consistent accretion and emission scenario in the 
system in the two observations, observed with a gap of $\sim$1 year.
\par
In EXO 0748$-$676, continuum of the eclipse spectra drops nearly by same factor throughout 
the (0.3-10) keV energy range from that of 
the OOE spectra in all the seven observations (Figure \ref{EXO_comp}). 
 The OOE to eclipse flux ratio has a short range (43-54) in the seven observations. These indicate
stability of X-ray reprocessing in the source during these observations.
All seven OOE spectra required partial covering absorption by partially ionized material for the best fit. 
The partially ionized materials cover ($55-92$$\%$) (Table \ref{EXO_OOE_zxipcf}) of the source emission. 
This component was not required to fit the eclipse spectra, which indicates that these ionized materials are either close
to the source and / or distributed in a narrow region, and fully blocked during eclipse.
 A weak blackbody emission (T$_{\rm BB}$: 0.09-0.21 keV) has been observed (with component \textit{bbodyrad}) during eclipse
 i.e. from a region at least at a distance equal to the radius of the companion from the neutron star.
 The size (radius) of these emitting regions are estimated to be 5-16 km using the blackbody normalization
 (($\frac{\rm R_{\rm km}}{\rm D_{10} \rm kpc})^{2}$: $123-325$, Table \ref{phenomeno}) and
 distance of the source (5.8$\pm$0.9 to 7.7$\pm$0.9 kpc, \citealt{Wolff_etal_2005}). 
 Also a weak black body emission (T$_{\rm in}$: 0.09-0.25 keV) from the disk is observed (with model component \textit{diskbb})
 during OOE phase of all the seven observations. The apparent inner accretion disk radius is estimated to be $320-3336$ km, from the 
 the disk normalization (($\frac{\rm R_{\rm in}}{\rm D_{10}})^{2}$$\rm cos\theta$: $860-23181$, 
 Table \ref{phenomeno}), distance to the source and inclination angle ((5-11) kpc, \citealt{2006Trigo}; 76.5$^{+1.4}_{-1.1}$, \citealt{Knight_etal2022}) of the source.
 A high temperature blackbody emission (T$_{BB}$: $\sim$4 keV) is observed in one observation (OB ID: 0160760101) during OOE phase. 
 The size (radius) of this blackbody emitting region is estimated to be ($0.68-2.58$ km), from the
 normalization (0.08$^{+0.03}_{-0.03}$) and distance to the source.
\par
 $1-5$ emission lines have been observed during eclipse phase of the seven observations of EXO 0748$-$676 (details are
 given in Table \ref{phenomeno_line}). The origin of 0.5 keV, 0.89 keV, 1.28 keV, 1.32 keV, 1.4 keV, 1.8 keV and 1.89 keV
 emission lines are possibly from O K$_{\alpha1}$, Ne K$_{\alpha1}$, Mg K$_{\alpha1}$, Mg K$_{\beta}$, Al K$_{\alpha1}$, Si
 K$_{\alpha1}$, Si K$_{\beta}$ respectively. $\sim$0.6 keV, $\sim$0.9 keV and $\sim$1.39 keV emission lines possibly are blend of
 other higher ionised emission lines with O K$_{\alpha1}$, Ne K$_{\alpha1}$ and Mg K$_{\beta}$ respectively.
 Detection of these lines in the eclipse phase indicates distribution of ionised Oxygen, Magnesium,
 Aluminium, Silicon in the extended medium surrounding the neutron star, which is at least at a distance equal to the radius of the
 companion from the neutron star. Emitting plasma consisting these ions may be present in the outer accretion disk or in the extended corona.
 Presence of $\sim$0.5 keV emission line during OOE phase in five observations indicates presence of
 ionised Oxygen close to the neutron star. Fe XXV / Fe XXVI absorption lines have been detected 
 during OOE phase in three observations (OB ID: 0160760401, 0160760601, 0160760801),
 these indicate that most of the photo-ionized absorbing plasma is located near the compact object.
Presence of extended corona have been proposed by \cite{2002Jimenez}.  In persistent spectra of  EXO 0748$-$676 with
 EPIC pn, \cite{2006Trigo} have found emission lines of O VII and Ne IX.
 \cite{2008Cottam} have observed Fe XXV, XXVI (n=2-3) and O VIII (n=1-2) absorption lines
 and emission lines from Neon, Nitrogen, Oxygen ions
 during bursts in EXO 0748$-$676 with \textit{XMM-Newton} RGS.  
 They propose photosphere of the neutron star as the origin of the 
absorption features and the
 emission lines possibly originate in the extended structure to the outer atmosphere.
\cite{2018Psaradaki}, with \textit{XMM-Newton} RGS detected several lines from ionized Neon, Nitrogen and Oxygen during eclipses. 
They confirm an extended atmosphere above the accretion disk.
\par
The composite eclipse spectra for the seven observations of this source almost overlap (top left panel of Figure \ref{EXO_comp}),
 this indicates nearly consistent spectra and reprocessed X-ray flux in eclipses
 during two months of observation period. This also 
 indicates similar kind of distribution of matter in the outer accretion disk and in the surrounding region, as during eclipse 
 inner accretion disk is blocked by the companion. Composite OOE spectra obtained including the dips intervals  of 
  the seven observations show an overlap above
 $\sim$4 keV while there is significant variation in the spectrum below 4 keV (top right panel of Figure \ref{EXO_comp}). 
 The variation in the low energy part of the spectrum is likely to be due to the variable patterns in the 
 inner accretion disk causing dips in EXO 0748$-$676 \citep{2017Gayathri}.
  The vertical structures that cause the dips in low energy X-rays, possibly vary in structure and density 
 in a manner, that they cause variation in the reprocessing of lower energy X-ray photons, whereas higher energy
 X-rays are almost unaffected by these structures. 
 The composite OOE spectra obtained excluding the dips intervals shows an overlap $\geq$2 keV (bottom panel).
 The variation $\leq$2 keV in the OOE spectra obtained excluding the dips intervals
 indicates some other structures/irregularities on the inner accretion disk 
 other than the vertical structures that casuses variation below 4 keV.
 Variation of soft X-rays could  also be due to different  accreting bulges 
 produced at the impact point between outer accretion disk and the accretion stream \citep{2017Gayathri}.
\par
 In AX J1745.6$-$2901, the line of sight equivalent Hydrogen column density (N$_{\rm H}$) found in both soft and hard state observations 
 during eclipse and OOE phases are comparable, 
 that indicates 
 nearly consistent environment in the system in both soft or hard states,
 during the period of the seven  observations.  
 N$_{\rm H}$ found during OOE and eclipse phase in AX J1745.6$-$2901 is nearly two
 orders of magnitude higher compared to that found
 in other eclipsing sources analysed in this text (Table \ref{phenomeno}). This indicates that the medium surrounding 
 the neutron star in this source is more dense compared to other three sources. 
The photon index of the eclipse spectrum is higher than that of 
the OOE spectrum in both the spectral states in AX J1745.6$-$2901, 
this indicates that the photons received during eclipses are not produced from pure
reflection of the primary source photons. We need a systematic analysis of OOE and eclipse 
spectra of the source with a large number of data sets to find out the exact cause of this behaviour.
\par
 Like MXB 1659$-$298, He like Fe XXV and H like Fe XXVI absorption 
line have been detected in the OOE phase in all the soft state obsevations of AX J1745.6$-$2901, which indicates hot, ionized material 
close to the neutron star during soft state.
No spectral line has been observed during OOE phase in the hard state.
\citealt{2015ponti_Fe}, \citealt{2017Ponti} have observed Fe K absorption  features  in the soft 
state with \textit{XMM-Newton} (\textit{XMM-Newton} and \textit{NuSTAR}), but these  features 
were not seen in the hard states spectra they have analyzed. Their \citep{2015ponti_Fe} analysis show 
% % %  that 
 the 
variability due to change in the  ionizing continuum, so they rule out over ionization 
as the reason for the disappearence of these absorption lines during hard state. 
They suggest that this behaviour could be due to the change of physical properties of plasma with the change of 
spectral energy distribution between the soft and hard state of the source.
\citet{2008_Hyodo}, had also found the same features in the soft state of the source 
with \textit{Suzaku}.  The disk corona could be the origin of these  absorption features (\citealt{2008_Hyodo},
\citealt{2015ponti_Fe}). \citet{2018Ponti} have obtained  highly variable absorption in the soft
states in the \textit{XMM-Newton} spectra while all the hard state spectra with \textit{XMM-Newton} and 
\textit{NuSTAR} were well fitted with constant column density of neutral absorption. They suggest that the  neutral  
absorption is mostly from the ISM.
\par
 We have observed high blackbody temperature ($\textgreater$ 1 keV) in the three soft state observations of AX J1745.6$-$2901 during OOE phase.
From the normalization 
(($\frac{\rm R_{\rm km}}{\rm D_{10} \rm kpc})^{2}$: $0.84-3.43$, Table \ref{phenomeno}) and distance to the source 
(8.3 kpc, \citealt{Genzel_etal_10}, \citealt{Bland_Gerhard_16}, \citealt{Gillessen_etal_2017}), the
size (radius) of the emitting regions is estimated to be 0.8-1.5 km.
In the soft-state below $\sim$3 keV the intensities in the eclipse and the OOE spectra 
differ by a lower factor ($\sim$3-5), than the intensities above $\sim$3 keV (rightmost figure of Figure \ref{AXJ_soft_spec}). 
But the hard state spectra (rightmost figure of Figure \ref{AXJ_hard_spec}) 
show that below 3 keV, there is negligible number of photons received during eclipse. 
These possibly indicates that the outer accretion disk is blocked during eclipses,
as most of the lower energy X-ray emission comes from the outer accreiton disk, which is not observed.
Soft-state composite OOE spectra of the system show nearly comparable
intensity for the three observations, while intensity for the other observations is nearly  
 more than 5 times lower than these three spectra (right figure of Figure \ref{AXJ_all_comp}). In the reprocessed
eclipse spectra also, the intensity of this soft state observation goes down by similar factor compared to the other three soft state observations
(left figure of Figure \ref{AXJ_all_comp}). These indicate nearly steady structure  of the accretion disk,
during these soft-state observations. OOE  to eclipse flux ratio
is comparable in the soft and hard states (Table \ref{phenomeno}), which indicates X-ray reprocessing property of the  matter surrounding
the compact object is independent of the spectral state of the source.
\par
EPIC pn eclipse spectrum of XTE J1710$-$281 has very poor statistics and the flux is very low
during this observation. The exposure time is 367 seconds for this observation, while the
source has an eclipse duration of 420 seconds \citep{2011Jain}. 
 With much less exposure time (98 seconds) we have observed more number of photons in one eclipse observation of MXB 1659$-$298 (OB ID: 0008620601).
The outer accretion disk radius in this sytem is found to be larger than 
the radius of the companion (R$_{\rm out}$ $\textgreater$ R$_{\rm C}$). 
The large OOE to eclipse flux ratio ($\sim$129) for this source perhaps indicates lower scale height of the
accreiton disk in this source. Because when the scale height is low, the solid angle covered by the outer accretion disk to the central source is low, which perhaps causes lesser reprocessing.
\par
To understand the types of occultation in the LMXB systems during maximum eclipse phase
(when the occultation is maximum in the whole eclipse phase) depending upon the binary separation $a$, radius of the companion R$_{\rm C}$, radius of the disk  R$_{\rm out}$ and the inclination $i$ of the system we have sketched three possibilities, which are given in Figure \ref{three_tpyes_occult}.
The values of R$_{\rm C}$, R$_{\rm out}$ and $a$ are listed in Table \ref{three_distances}.
\begin{table*}[ht]
% %  \tiny
% \begin{center}
\label{table}
\vspace{1ex}
\caption{List of radius of the companion star R$_{\rm C}$, radius of the accreiton disk
R$_{\rm out}$ and the orbital separation $a$ respectively for the four LMXB sytems.}
\centering
% \begin{small}
\begin{tabular}{l rrr}
\hline

  Source	       & R$_{\rm C}$            & R$_{\rm out}$      & $a$    \\

                   & (10$^{8}$ m)           & (10$^{8}$ m)       & (10$^{8}$ m)   \\
%   	               &                        &                   &      \\

		\hline
% 		  	               &                        &                   &      \\
MXB 1659$-$298	   & (5.15-6.54)$^{a_{2}}$     & (3.75-9.19)\textcolor{magenta}{$^\diamondsuit$}       & (15.47-19.31)\textcolor{magenta}{$^\spadesuit$}   \\
%   	               &                         &                   &      \\
		\hline
EXO 0748$-$676	   & 2.992$^{b_{2}}$           &  (3.31-6.87)\textcolor{magenta}{$^\diamondsuit$}      & (10.88-12.18)\textcolor{magenta}{$^\spadesuit$}  \\
%   	               &                         &                   &      \\
		\hline
AX J1745.6$-$2901 &  6.40\textcolor{magenta}{$^{\star}$}           &  6.39\textcolor{magenta}{$^\diamondsuit$}             &  18.85\textcolor{magenta}{$^\spadesuit$} \\
%   	               &                         &                   &      \\
		\hline
XTE J1710$-$281   & 2.51\textcolor{magenta}{$^{\star}$}            &  3.66\textcolor{magenta}{$^\diamondsuit$}             & 9.43\textcolor{magenta}{$^\spadesuit$}  \\
%   	               &                         &                   &      \\
		\hline
%   	               &                         &                   &      \\

 \end{tabular}
\label{three_distances} \\
\begin{tablenotes}
 \item[*] 
\small{$^{a_{2}}$\citep{2018Iaria}, $^{b_{2}}$\citep{Knight_etal2022}} \\
\textcolor{magenta}{$^{\star}$}Derived from orbital period ($\frac{\rm R_{\rm C}}{\rm R_{\odot}}$ $\sim$0.11P$_{\rm orb}$, \citealt{Frank_etal_02}) \\
\textcolor{magenta}{$^\diamondsuit$}Estimated using the relation R$_{\rm out}$ $\sim$0.8$\times$R$_{\rm LNS}$, where R$_{\rm LNS}$ is the Roche lobe radius of the neutron star, details are given in section 1. \\
\textcolor{magenta}{$^\spadesuit$}Calculated using Keplar's third law, as mentioned in section 1)\\
\end{tablenotes}

\end{table*}
% % %
% ,
%
The projection of the circular accretion disk of radius R$_{\rm out}$ on the
sky plane (the plane perpendicular to the line of sight) is an ellipse of semi major
axis length R$_{\rm out}$ and semi minor axis length R$_{\rm out}$$\cos(i)$.
The projection of spherical companion star of radius R$_{\rm C}$ on the sky
 plane is a circle of radius R$_{\rm C}$.
 The projection of the companion star may fully cover the projection of the accretion disk, or partially cover the projection of the accretion disk with two or four intersections (left, middle, right figure respectively in Figure \ref{three_tpyes_occult}). The four point intersection is given in \cite{Court_etal2019}.
 Assuming a small tilt of the accretion disk with the orbital plane (as the tilt is not known accurately) and no wrap in the disk, we have derived the occulted area of the projection of the accretion disk on the sky plane by the projection of the companion star during maximum eclipse phase.
 In the first case (when the accretion disk is fully covered by the companion star) the occulted area is the full area
of the ellipse on the sky plane.
In the other two cases first we obtained the  intersection points by solving equations of ellipse and circle. Then integrating over the overlapping area
of the ellipse and the circle between the intersection points we obtained the total occulted area of the disk. The details of the derivations of the occulted and hence visible area of the accretion disk are given in the Supplementary Materials in Section 5.
\par
To understand the variation of the visibility of the accreiton disk as functions of orbital
parameters and the radius of the companion and the accreiton disk, in Figure \ref{June15_Vis_six_panel} we have plotted these
parameters against the percentage of the accretion disk that is visible during total eclipse. The expression for the visibility of the accreiton disk ($\%$) can be found in the Supplementary Materials in section 5. It is obtained by dividing the visibile area of the disk by total area of the disk multiplied by 100. It is a function of the inclination of the system $i$, radius
of the disk R$_{\rm out}$ and radius of  the companion star  R$_{\rm C}$ and binary separation $a$. As the binary separation $a$ and the orbital period  P$_{\rm orb}$ is correlated (the first equation given in section 1), the visibility is also a function of  P$_{\rm orb}$.
Mass ratio  $q$ between the companion and the neutron star and binary separation  $a$ are also correlated (3rd equation of section 1), so the visibility of the disk is also a function of $q$.
The variation of visibility with  $a$ and P$_{\rm orb}$ are similar, which is in concilliation with equation 1.
We see high anti-correlation between the visibility of the disk and both $a$ and   P$_{\rm orb}$ with correlation coefficient ($r$) of -0.84 and -0.89 respectively (with the visibility).
 R$_{\rm C}$  is also highly anticorrelated with the visibility with $r$ = -0.85.
 $q$ and R$_{\rm out}$ are moderately anticorrelated with the visibility with $r$ of -0.72 and
-0.67 respectively.
A positive correlation is found between the inclination $i$ and the visibility of the disk
with $r$ = 0.90 and R$^{2}$ (goodness of fit) = 0.80.
The occultation is observed to be more when i$\sim$70° compared to the case when $i$ is $\sim$80°.
\par
To understand the XRP efficiency (reciprocal of OOE to eclipse flux) as functions of binary characteristics,
we have plotted the orbital parameters (P$_{\rm orb}$, $q$, $i$, $a$), radius of the companion R$_{\rm C}$ and mass accretion rate $\dot{\rm M}$ with the variation of OOE to eclipse flux ratio in Figure \ref{June15_fratio_five_panel}.
To estimate  $\dot{\rm M}$ we have obtained
 total unabsorbed flux (F) in 3-10 keV energy range for each observation during OOE phase (using {\small  CFLUX}
 command in {\small  XSPEC}) and used the expression $\eta$$\times$dM/dt$\times$c$^{2}$ = F$\times$4$\pi$d$^{2}$,
 where $\eta$ is the mass to energy conversion efficiency in a neutron star (0.15, \citealt{Frank_etal_02}),
 c is speed of light in free space and d is the distance to the source.
 P$_{\rm orb}$ and R$_{\rm C}$ are observed to be highly anticorrelated with the OOE to eclipse flux ratio with r of -0.83 and -0.83 respectively (with the flux ratio). $q$ is moderately anticorrelated with the OOE to eclipse flux ratio with $r$ = -0.71. $i$ is highly correlated with the flux ratio with $r$ = 0.83.
Very weak anti-correlation is seen between the mass accretion rate $\dot{\rm M}$ and the flux ratio with $r$ = -0.15.
% % %
%
\par
In Figure \ref{June15_Vis_XRP} we have plotted the XRP effeciency with the percentage of the projected disk area that is visible during maximum eclipse phase (visibility of the disk in $\%$). A correlation is expected between the X-ray reprocessing efficiency and the visible area of the accrtion disk during maximum eclipse. But this is not observed, rather a negative relation is observed with $r$ = -0.99 and R$^{2}$ = 0.99.
The reason of which is not clear from this small sample size and with large uncertainties.
\par
The uncertainties in the orbital parameters lead to the uncertainties in the estimated parameters and hence to the visibility of the disk.
We need a large sample and more accurate orbital parameters to find out the exact relationship between the orbital parameters (and also the estimated parameters), the visibility of the disk and the X-ray reprocessing efficiencies of the eclipsing systems.
\par
The system to system differences and unexpected relations between few system parameters with the
X-ray reprocessing efficiencies indicate that the X-ray reprocessing characteristics are very sensitive to the
detailed geometries of the systems, some of which (scale height of the accretion disk, tilt angle of the accretion disk with respect to the binary orbit/line of sight etc.) are not accurately measurable till date. 
\par
\subsection{\underline{SUMMARY}}
We summarize the results as follows:
Flux ratio of the eclipse to OOE spectra shows a range
from $\sim$6 (in AX J1745.6$-$2901, OB ID: 0723410501) to $\sim$129 (in XTE J1710$-$281, OB ID: 0206990401), 
which differs by a factor of $\sim$21. 
This flux ratio is nearly same in both soft and hard states in AX J1745.6$-$2901.
No spectral line has been found in AX J1745.6$-$2901 in any of the phase during hard states.  Three sources 
(MXB 1659$-$298,  EXO 0748$-$676, AX J1745.6$-$2901 in soft state) 
show absorption lines during OOE phases. 1-5 emission lines have been observed during eclipse
phase of the seven observations of EXO 0748$-$676.
 Line of sight equivalent Hydrogen 
column density is found to be much higher in  AX J1745.6$-$2901 compared to other sources.
During out-of-eclipse, in EXO 0748$-$676, the compact object is found to be partially 
covered by partially ionized material by at least 55 $\%$.
Comparison of this various aspects of the X-ray reprocessing suggests significant differences
in the X-ray reprocessing in these LMXBs.  The differences could be due to could be due to differences in the 
 scale height of the accretion disk, relative size of the  accretion disk with respect to the companion, binary separation, mass ratio between the neutron star and the companion, mass accretion rate extension of the accretion disk atmosphere/wind, composition
of the scattering agents, their density, ionization level, distribution etc. in these LMXBs,
despite all being dipping and eclipsing. 
\par
\subsection{\underline{COMPARISON OF X-RAY REPROCESSING BETWEEN HMXB AND LMXB SYSTEMS}}
In case of HMXBs we have seen much wider variation  in the OOE to eclipse flux ratio ($\sim$8-237),
which is perhaps due to differences in the stellar wind mass loss properties of the HMXB companions \citep{Aftab2019}.
In seven observations of AX J1745.6$-$2901 and EXO 0748$-$676, the OOE to eclipse flux ratio have been
found to have a narrow range, ($\sim$6-10) and ($\sim$43-54)
respectively. Two observations of MXB 1659$-$298 also show similar  OOE to eclipse flux ratio (21-24).
These indicate less variable environment surrounding disk-fed systems.
These flux ratios are smaller factor compared to wind accreting
supergiant HMXBs \citep{Aftab2019},
where we have seen large variation even in the same source  (as much as a factor of $\sim$3 in LMC X$-$4) within a 
period of less than a year. 
The smaller value of the OOE to eclipse flux ratio observed in LMXBs indicates greater  reprocessing in LMXBs,
even though the medium of LMXBs are less dense and less extended compared to that in HMXBs.
 In the eclipse spectra of HMXBs, 6.4 keV Fe emission line was very common, whereas this line is hardly seen
in the eclipse spectra of the LMXBs.
These are important outcomes of the two work: Xray reprocessing in HMXBs and LMXBs.
Further analysis of a large number of data sets from HMXB and LMXB systems can get us clear picture about the
X-ray reprocessing effeciency and its dependencies in the two types of systems.\\
\section{Conclusion} 
- Despite all the sources being dipping and eclipsing, comparison of various aspects of the X-ray
 reprocessing suggests significant differences in the X-ray reprocessing in these LMXBs, 
 which could be due to differences in the inclination angle, scale height of the accretion disk, relative size of the  accretion disk with respect to the companion, binary separation, mass ratio between the neutron star and the companion, mass accretion rate, composition of the scattering agents, their density, ionization level, distribution etc.\\ \\
- This study presents, for the first time, the derivation of the occulted area of the accretion disk by the companion star (and consequently, the visible area of the disk) for various obscuration geometries during the maximum eclipse phase. \\ \\
- The eclipse spectra for some of the LMXB sources are being reported for the first time. \\ \\
- The primary X-ray source and the eclipse spectrum
 remains nearly same over two months in EXO 0748$-$676. Different types of vertical structures on the accretion disk
 produce variation in the lower energy X-rays in the OOE spectrum. \\ \\
- OOE to eclipse flux ratio are observed to be comparable, irrespective of the intensity state in  AX J1745.6$-$2901. \\ \\
- Large OOE to eclipse flux ratio and hence low X-ray reprocessing efiiciency in XTE J1710$-$281 indicates low scale height in this system. \\ \\
- Despite less extended and less dense environment in LMXBs compared to HMXBs, more X-ray reprocessing is observed
 in the LMXBs.\\ \\
- Most contrasting outcome with respect to the HMXBs is the lack of strong 6.4 keV 
 line in the eclipse spectrum of LMXBs. \\ \\
% % % =============================================================================================================================================
% 
\par
\textbf{ACKNOWLEDGEMENT}\\
The data used for this work has been obtained through the High Energy Astrophysics Science Archive (HEASARC) On-
line Service provided by NASA/GSFC.  We thank Dr. Wasim Aftab for his help to automate the
 {\it XMM-Newton} EPIC-pn script and  Anisa Aftab for her co-operation in solving different technical issues during this work.  NA would like to thank Dr. Pragati Pradhan and Dr. Gayathri Raman for the constructive discussions related to pn data analysis. NA  would like to express sincere thanks to Dr. Dhruv Pathak for the valuable discussions on the theoreticaal facet of eclipsing systems.
\\
\\
% ---------------------------------------
\par
\section{\textbf{Supplementary Materials}}
% \\
% \\
In section 3, we have seen that depending upon Radius of the accretion disk R$_{\rm out}$. radius of the companion R$_{\rm C}$, binary separation $a$ and inclination $i$ three scenario can happen during maximum eclipse phase. Figure \ref{three_tpyes_occult} describes the three scenario.
In the first case occulted disk area is 100$\%$, so the visibile area of the disk is 0$\%$.
Below we show the derivations of the occulted area and hence visibile area of the disk during maximum eclipse phase in terms of the above parameters.\\ \\
(i) \textbf{Derivation of the occulted and hence the visibile area of the projection of the accretion disk during maximum eclipse phase, when the projection of the companion intersects the projected disk at two ponts:}\\
\\
% ---------------------------------------------------
%
\begin{figure*}
\centering
% % % \includegraphics[scale=0.8, angle=90]{latest_2p.eps}
\includegraphics[scale=0.8, angle=90]{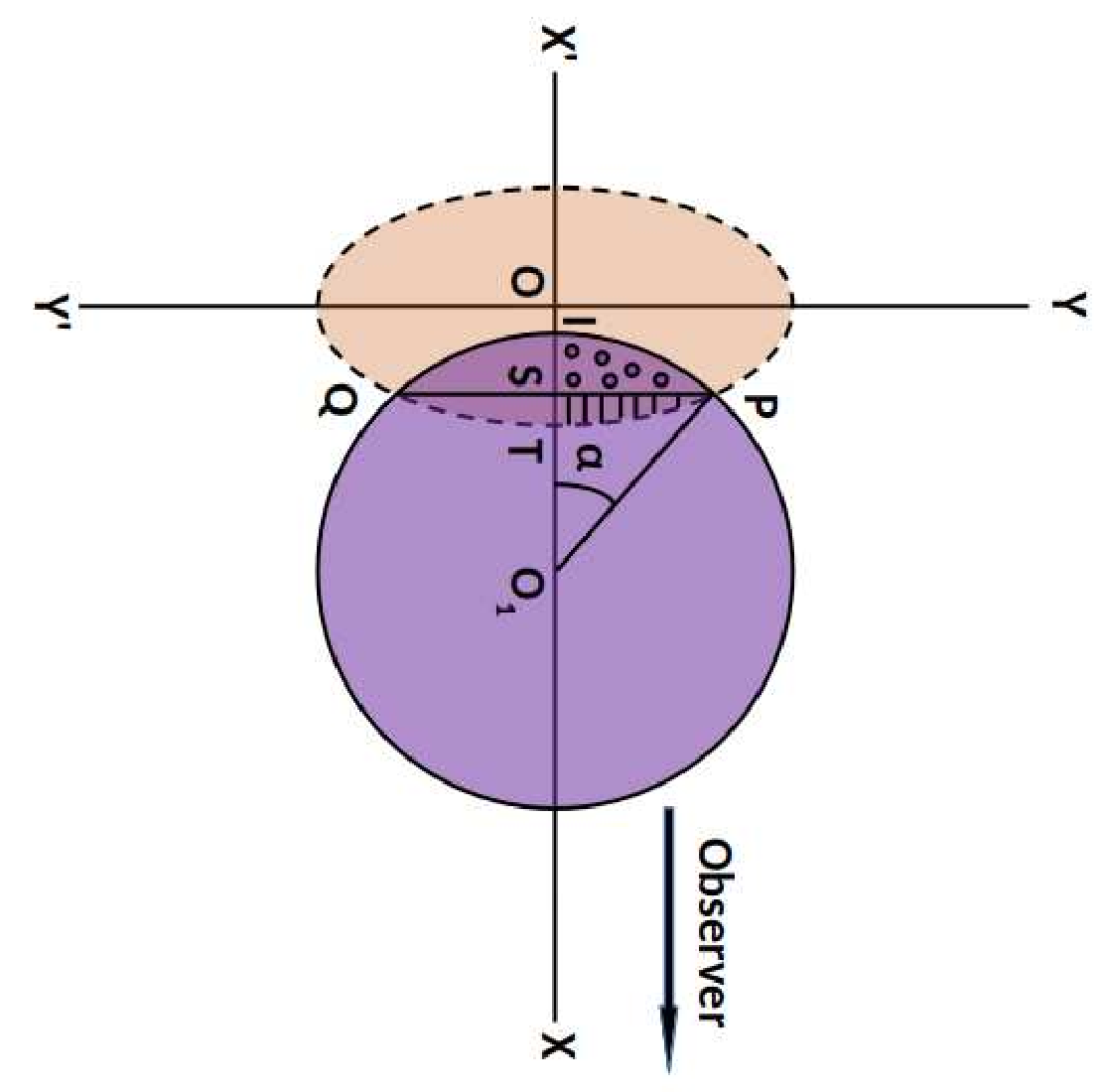}
\caption{\large{The projection of the accretion disk on the sky plane (ellipse) is intersected by the projection of the companion star at two points (P and Q). O and O1 are the centers of the ellipse and the circle respectively. The circle and the ellipse intersect the X axis at I and T respectively. The occulted area of the accretion disk is the area covered within PIQTP. The area within PITP is symmetrical to the area covered by QITQ. Hence the total occulted area (within PIQTP) is double the area covered by PITP. The area within PISP (shown with small hollow circles) is the area covered by the circular arc PI and the straight lines IS and SP. The area within PSTP (shown with small vertical bars) is the area covered by the elliptic arc PT and the straight lines ST and SP.
The visible area of the accretion disk is obtained subtracting the occulted area from the total area  of the ellipse. The derivation of the occulted area and hence the visible area given in section 5.}}
\label{details_2pt_intersect}
\end{figure*}
% ---------------------------------------
In Figure \ref{details_2pt_intersect} we see the projection of the accretion disk on the sky plane is intersected by the projection of the companion star at two points (P and Q). The inclination of the system is $i$, i.e. normal to the orbit of the binary makes an angle $i$ with the line of sight (LOS).
The projection of the circular accretion disk of radius $\rm R_{out}$ on the sky plane (the plane perpendicular to the LOS) is an ellipse of semi-major axis $\rm R_{out}$ and semi-minor axis
$\rm R_{out}$$\cos$(i). The projection of the spherical companion of radius
 $\rm R_{C}$ on the sky plane is a circle of radius $\rm R_{C}$.
The centre of the projected  ellipse and the circle are respectively O and O$_{1}$ on the sky plane. We have chosen the origin at O. The  X axis is along the line joining O and O$_{1}$. OO$_{1}$ = $a\cos (i)$, as $a$ is the binary separation.
The whole occulted area is the area enclosed within circular arc PIQ and elliptic arc PTQ. The circular arc PI makes an angle $\alpha$ at O$_{1}$.
The X axis divides this area symmetrically into two halves (enclosed within PITP and QTIQ). So the total occulted area is double the area within PITP. This area is the sum of the area enclosed by circular arc PI, straight line IS, SP (within PISP) and the area enclosed by elliptic arc PT, straight lines TS and SP (within PSTP). These two areas (say A$_{c}$ and A$_{el}$ respectively) are shown in Figure \ref{details_2pt_intersect} with hollow circles and vertical bars respectively.
\par
The area A$_{c}$ can be obtained by subtracting the area of the triangle PSO$_{1}$ (say A$_{4}$) from
the area of the circular arc PI  made at its centre O$_{1}$ (say A$_{3}$), i.e. A$_{\rm c}$ = A$_{3}$ - A$_{4}$.
Solving the equation of the circle ((x$_{1} - a\cos (i))^{2}$  + y$^{2}$ = R$_{C}$$^{2}$ [Eq. 1]) and the ellipse (x$^{2}$/a$_{mj}$$^{2}$ + y$^{2}$/a$_{mn}$$^{2}$ = 1 [Eq. 2]) we obtained the abscissa (x$_{1}$) of point P (x$_{1}$, y$_{1}$). where The semi major axis length of the ellipse = a$_{\rm mj}$ = $\rm R_{out}$ and the semi minor axis length of the ellipse = a$_{mn}$ = \rm R$_{\rm  out}$$\cos$(i).\\
$y_{1} = \rm R_{C}\sin \alpha$ (from Figure \ref{details_2pt_intersect})\\
$ x_{1} = \frac{-2\rm a \cos i  + \sqrt{ 4\rm a^{2}\cos^{2}i - 4\tan^{2}i(\rm R_{C}^2 - \rm R_{\rm out}^2 - a^{2}\cos^{2}i)}}{2\tan^2i} $ (solving Eq. 1 and Eq. 2) \\
From the equation of the circle (Eq. 1) with x = x$_{1}$, y = y$_{1}$ = $\rm R_{C}\sin \alpha$
we get
$\alpha = \sin ^{-1}\frac{\sqrt{\rm R_{C}^2 - (x_{1} - a\cos (i))^{2}}}{\rm R_{C}}$ \\
% -------------
% -------------
$A_{3} =  \frac{1}{2}R_{C}^{2}\alpha$ (from Figure \ref{details_2pt_intersect})\\
$A_{4} =   \frac{1}{4}\rm R_{C}^{2}\sin 2\alpha$ (from Figure \ref{details_2pt_intersect})          \\
Hence, A$_{\rm c}$ = A$_{3}$ - A$_{4}$ =  $\frac{1}{4}\rm R_{C}^{2}(2\alpha - \sin 2\alpha)$\\
$A_{\rm el} = \int_{x_{\rm S}}^{x_{\rm T}} y \, dx$, where x$_{\rm S}$ = x$_{1}$ and
x$_{\rm T}$ =  R$_{out}$ $\cos$(i)\\
$A_{\rm el} = a_{mn}\int_{x_{\rm 1}}^{R_{\ out} \cos(i)} \sqrt{(1-\frac{x^2}{a_{mj}^2})} \, dx$ \\
 Inegratiing we get,
$A_{\rm el} = \frac{\rm R_{\rm out}^2\cos(i)}{4}(\pi - 2\theta - \sin 2\theta)$,
where $\theta = \sin^{-1}\frac{x_{1}}{\rm R_{out}\cos(i)}$\\
Hence, the total occulted area of the accretion disk (A$_{\rm occult, 2p}$) = 2(A$_{c}$ + A$_{el}$) \\
The area of the accreiton disk = A$_{\rm disk}$ = $\pi$a$_{mj}$a$_{mn}$ = $\pi$R$_{\rm out}$$^{2}$cos(i)  \\
The visible area of the accretion disk = A$_{\rm visible, 2p}$ = A$_{\rm disk}$ - A$_{\rm occult, 2p}$\\
The perecentage of the area of the accreiton disk visible during total eclipse = A$_{\rm visible, 2p}$ ($\%$) =
(A$_{\rm visible, 2p}$/A$_{\rm disk}$)$\times$100$\%$
\\
\\
(i) \textbf{The derivation of the occulted area of the projection of the accretion disk during total eclipse when the projection of the companion intersects it at four ponts:}\\
\\
  \begin{figure*}
\centering
% % % \includegraphics[scale=0.8, angle=90]{latest_4p.eps}
\includegraphics[scale=0.8, angle=90]{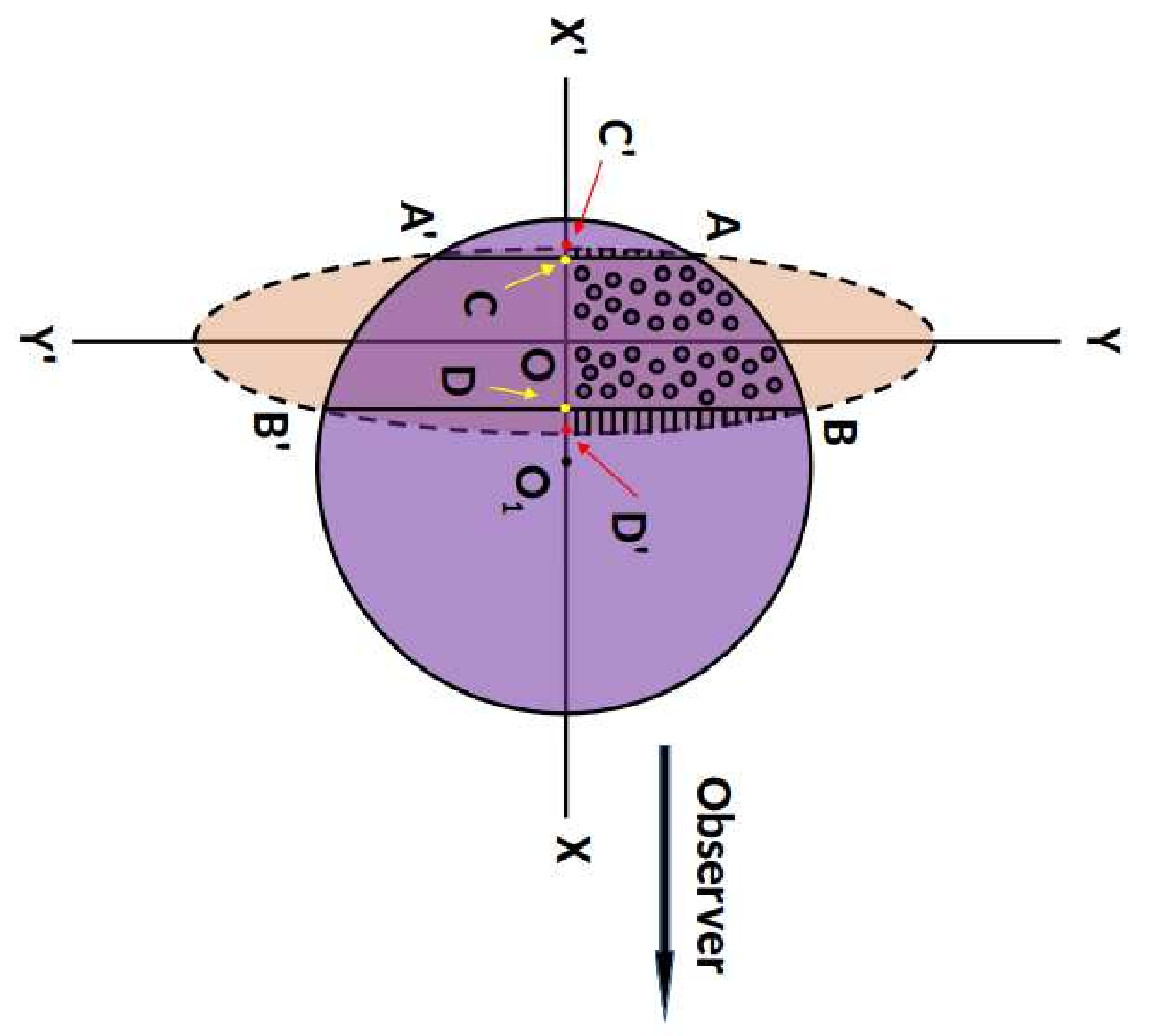}
\caption{Similar to Figure \ref{details_2pt_intersect}. Here the projection  of the accretion disk on the sky plane (ellipse) is intersected by the projection of the companion star at four points (A, B, A$'$ and B$'$). The straight line AA$'$ and BB$'$ intersect the X axis at point C and point D respectively. The ellipse intersects the X axis at two points C$'$ and D$'$. The occulted area during maximum eclipse is the area covered by the circular arc AB and A$'$B$'$ and elliptic arc AA$'$ and BB$'$  (i.e. within AA$'$B$'$BA). This area is double the area covered by ABD$'$C$'$A. The region filled with hollow circles is the area covered by circular arc AB and straight lines BD, DC and CA. The two regions filled with vertical bars are covered by AC$'$CA and BDD$'$B are the regions enclosed by the elliptic arc AC$'$, straight lines C$'$C, CA and elliptic arc BD$'$, straight lines D$'$D, DB respectively. The derivation of these areas and finally the visible area are given in section 5.
}
\label{details_4pt_intersect}
\end{figure*}
  % ---------------------------------------
Figure \ref{details_4pt_intersect} shows that the projection of the accretion disk on the sky plane is intersected by the projection of the companion star at four points (A, B, A$'$, B$'$). The description of the geometry remains same as Figure \ref{details_2pt_intersect}. The occulted area of the projection of the accretion disk is the area enclosed within the circular arcs AB, A$'$B$'$ and elliptic arcs AA$'$, BB$'$. The line joining O and O$_{1}$ i.e. the X axis divides this area symmetrically into two halves (area within ABD$'$C$'$A and A$'$B$'$D$'$C$'$A$'$). The area enclosed by the circular arc AB and the straight lines BD, DC and CA are shown with hollow circles (within ABDCA, say A$_{\rm C1}$). The area enclosed by the elliptic arc C$'$A, straight lines AC, CC$'$ (within AC$'$CA, say A$_{\rm el1}$) and that within elliptic arc BD$'$, straight lines D$'$D, DB  (within D$'$BDD$'$, say A$_{\rm el2}$) are shown with vertical bars.
Solving the equation of the circle (Eq. 1) and the ellipse (Eq. 2) we obtain the abscissas (x$_{2}$, x$_{3}$) of the point A (x$_{2}$, y$_{2}$) and B (x$_{3}$, y$_{3}$).\\
$x_{3}, x_{2} = \frac{-2\rm a \cos i  \pm \sqrt{ 4\rm a^{2}\cos^{2}i - 4\tan^{2}i(\rm R_{C}^2 - \rm R_{\rm out}^2 - a^{2}\cos^{2}i)}}{2\tan^2i}$\\
$A_{\rm el1} =  \int_{x_{\rm C'}}^{x_{\rm C}} y \, dx$, where x$_{\rm C'} = -\rm R_{out}\cos(i)$,
x$_{\rm C} = x_{2}$ \\
% $A_{\rm el1} =  a_{mn}\int_{x_{\rm C'}}^{x_{\rm C}} \sqrt{(1-\frac{x^2}{a_{mj}^2}) \, dx$
$A_{\rm el1} = a_{mn}\int_{-\rm R_{\ out} \cos(i)}^{x_{\rm 2}} \sqrt{(1-\frac{x^2}{a_{mj}^2})} \, dx$ \\
Deriving the above integration we get,\\
$A_{\rm el1} = \frac{\rm R_{out}^2\cos(i)}{4}(\pi + 2\theta_{2} + \sin2\theta_{2})$ where $\theta_{2} = \sin^{-1}\frac{x_{2}}{\rm R_{out}\cos(i)} $\\
$A_{\rm el2} =  \int_{x_{\rm D}}^{x_{\rm D'}} y \, dx$, where x$_{\rm D} = x_{3}$ and
x$_{\rm D'} = \rm R_{out}\cos(i)$ \\
$A_{\rm el2} = a_{mn}\int_{x_{\rm 3}}^{\rm R_{\ out} \cos(i)} \sqrt{(1-\frac{x^2}{a_{mj}^2})} \, dx$ \\
$A_{\rm el2} = \frac{\rm R_{out}^2\cos(i)}{4}(\pi - 2\theta_{3} - \sin2\theta_{3}) $ where
$\theta_{3} = \sin^{-1}\frac{x_{3}}{\rm R_{out}\cos(i)} $\\  \\
$A_{\rm C1} =  \int_{x_{\rm C}}^{x_{\rm D}} y \, dx$, where x$_{\rm C} = x_{2}$ and
 x$_{\rm D} = x_{3}$ \\
$A_{\rm C1} =  \int_{x_{2}}^{x_{3}} \sqrt{\rm R_{C}^2 - (x - a\cos(i))^2} \, dx$ \\
Deriving the above integration we get,\\
$A_{\rm C1} = \frac{\rm R_{C}^2}{4}[2(\theta_{max} - \theta_{min}) + \sin2\theta_{max} - \sin2\theta_{min}]$\\
where $\theta_{max} = \sin^{-1}\frac{x_{3} - a\cos(i)}{\rm R_{C}}$
and $\theta_{min} = \sin^{-1}\frac{x_{2} - a\cos(i)}{\rm R_{C}}$ \\
Hence, the total occulted area of the accretion disk (A$_{\rm occult, 4p}$) =
2(A$_{\rm el1}$ + A$_{\rm el2}$ + A$_{\rm C1}$) \\
The visible area of the accretion disk = A$_{\rm visible, 4p}$ = A$_{\rm disk}$ - A$_{\rm occult, 4p}$ \\
The perecentage of the area of the accreiton disk visible during total eclipse = A$_{\rm visible, 4p}$ ($\%$)
(A$_{\rm visible, 4p}$/A$_{\rm disk}$)$\times$100$\%$

\begin{thebibliography}{00}
\bibitem[Aftab, Paul and Kretschmar(2019)]{Aftab2019} Aftab N., Paul B., Kretschmar P. X-Ray Reprocessing: Through the Eclipse Spectra of High-mass X-Ray Binaries with \textit{XMM-Newton}. ApJS  2019; 243: 29.

\bibitem[Arnaud (1996)]{Arnaud1996} Arnaud, K. A.XSPEC: The First Ten Years.  ASP Conf. Series, Vol. 101, Astronomical Data Analysis Software and Systems, ed. G. H. Jacoby $\&$ J. Barnes, 17. 

\bibitem[Bland et~al.(2016)]{Bland_Gerhard_16} Bland-Hawthorn and Gerhard. The Galaxy in Context: Structural, Kinematic, and Integrated Properties.  Annu. Rev. Astron. Astrophys 2016; 54: 529-596. 

\bibitem[Boirin et~al.(2007)]{2007Boirin}Boirin L., Keek L., Méndez M., Cumming A., in’t Zand J. J. M.,
Cottam J., Paerels F., Lewin W. H. G. Discovery of X-ray burst triplets in EXO 0748$-$67.  A$\&$A 2007; 465: 559-573.

\bibitem[Cominsky et~al.(1984)]{1984Cominsky}Cominsky L. R., Wood K. S. Discovery of a 7.1 hour period and eclipses from MXB 1659$-$29. ApJ 1984; 283: 765-773. 
%
\bibitem[Cominsky et~al.(1989)]{1989Cominsky}Cominsky L. R., Wood K. S. Further Observations of the Eclipsing X-Ray Burst Source MXB 1659$-$29. ApJ 1989; 337, 485.
%
\bibitem[Cottam et~al.(2008)]{2008Cottam}Cottam J., Paerels F., Méndez M., Boirin L., Lewin W. H. G.,
Kuulkers E., Miller J. M. The Burst Spectra of EXO 0748$-$676 during a Long 2003 \textit{XMM-Newton} Observation.  ApJ 2008; 672: 504-509. 
%
\bibitem[Court et~al.(2019)]{Court_etal2019}Court et al. 2019. The eclipsing accreting white dwarf Z chameleontis as seen with TESS. MNRAS, MNRAS 2019, 488, 4149-4160
%
\bibitem[de Jong et~al.(1996)]{1996Jong}de Jong, J. A., van Paradijs, J., $\&$ Augusteijn, T. Reprocessing of X rays in low-mass X-ray binaries. A$\&$A 1996; 314, 484-490
% 
\bibitem[Trigo and Boirin (2016)]{Trigo_Boirin_16}Dıaz Trigo M. $\&$ Boirin L.  
 Accretion disc atmospheres and winds in low-mass X-ray binaries. Astron. Nachr. 2016; 337, 368
% 
%
\bibitem[Trigo et~al.(2011)]{2011Trigo_0748}Diaz Trigo M., Boirin L., Costantini E., Méndez M., Parmar A. \textit{XMM-Newton}
observations of the low-mass X-ray binary EXO 0748$-$676 in quiescenc.  A$\&$A 2011; 528, A150.
%
\bibitem[Trigo et~al.(2006)]{2006Trigo}Diaz Trigo M., Parmar A. N., Boirin L., Mendez M., Kaastra J. S. Spectral changes during dipping in low-mass X-ray binaries
due to highly-ionized absorbers. A$\&$A 2006; 445: 179-195.

\bibitem[Frank, King and Lasota(1987)]{Frank_etal_87} Frank, J., King, A. and Lasota J. The light curves of Low Mass X-ray Binaries. A$\&$A 1987; 178: 137-142.

\bibitem[Frank, King and Raine(2002)]{Frank_etal_02}Frank, J., King, A. and Raine, D. Frank, J., King, A. and Raine, D.  Accretion Power in
Astrophysics. Cambridge University Press, Cambridge. 2002.
% % % 
% % % 
\bibitem[Galloway et~al.(2008)]{Galloway2008}Galloway D. K., Muno M. P., Hartman J. M., Psaltis D., Chakrabarty D. Thermonuclear (Type I) X-Ray Bursts Observed by the \textit{Rossi X-Ray Timing Explorer}.  ApJS 2008; 179: 360-422.
% % % 
% % % 
\bibitem[Genzel et~al.(2010)]{Genzel_etal_10}Genzel R. et al. A study of the gas-star formation relation over cosmic time. MNRAS 2010; 407: 2091-2108.
% % % 
% % % 
\bibitem[Gierli{\'n}ski et~al.(2009)]{2009Gierlinski}Gierli{\'n}ski, M., Done, C., $\&$ Page, K. Reprocessing of X-rays in the outer accretion disc of the black hole binary XTE J1817$-$330. MNRAS 2009; 392, 1106-1114
% % % 
% % % 
\bibitem[Gillessen et~al.(2017)]{Gillessen_etal_2017}Gillessen S. et al.
An Update on Monitoring Stellar Orbits in the Galactic Center. ApJ 2017; 837, 30.% % % 
% % % 
\bibitem[Hynes and Jones (2009)]{2009Hynes}Hynes R. I., Jones E. D. The Quiescent Optical and Infrared 
Counterpart to EXO 0748$-$676 = UY Vol.  ApJL 2009; 697: L14-L17.
% 
\bibitem[Hyodo et~al.(2009)]{2008_Hyodo}Hyodo Y., Ueda Y., Yuasa T., Maeda Y., Makishima K., Koyama K. Timing and 
Spectral Study of AXJ1745.6$-$2901 with \textit{Suzaku}. PASJ 2009; 61: S99-S106.
% % % 
% % % 
\bibitem[Iaria et~al.(2018)]{2018Iaria}Iaria R. et al.
A possible solution of the puzzling variation of the orbital period of MXB 1659-298. MNRAS 2018; 473: 3490-3499.
% % % 
% % % 
\bibitem[Jain  and Paul (2011)]{2011Jain}Jain C., Paul B.
Eclipse timings of the LMXB XTE J1710$-$281: orbital period glitches. MNRAS 2011; 413: 2-6.
% % % 
% % % 
\bibitem[Jain et~al.(2017)]{2017Jain_1659}Jain C., Paul B., Sharma R., Jaleel A.
Indication of a massive circumbinary planet orbiting the low-mass X-ray binary MXB 1658-298. MNRAS 2017; 468: L118-L122.
% % % 
% % % 
\bibitem[Jain et~al.(2022)]{2022Jain_1710}Jain C., Sharma R., Paul B.
Eclipse Timings of the LMXB XTE J1710-281 : discovery of a third orbital period glitch. MNRAS 2022; 517: 2131-2137.
% % % 
% % % 
\bibitem[Jimenez et~al.(2002)]{2002Jimenez}Jimenez-Garate M. A., Raymond J. C., Liedahl D. A.
The Structure and X-Ray Recombination Emission of a Centrally Illuminated Accretion Disk Atmosphere and Corona. ApJ 2002; 581: 1297-1327.
% % % 
% % % 
\bibitem[Jin et~al.(2017)]{2017Jin}Jin C., Ponti G., Haberl F., Smith R.
Probing the interstellar dust towards the Galactic Centre: dust-scattering halo around AX J1745.6$-$2901. MNRAS 2017; 468: 2532-2551.
% % % 
% % % 
\bibitem[Jin et~al.(2018)]{2018Jin}Jin C., Ponti G., Haberl F., Smith R., Valencic L.
Effects of interstellar dust scattering on the X-ray eclipses of the LMXB AX J1745.6$-$2901 in the Galactic Centre.  MNRAS 2018; 477: 3480-3506.
% % % 
% % %
\bibitem[Knight et~al.(2022)]{Knight_etal2022}Knight Amy H., Ingram Adam, Middleton Matthew and Drake Jeremy.
Eclipse mapping of EXO 0748-676: evidence for a massive neutron star. MNRAS 2022; 510: 4736-4756.
% % % 
% % % 
\bibitem[Kubota et~al.(1998)]{Kubota_etal_98}Kubota A. et al.
Evidence for a Black Hole in the X-Ray Transient GRS 1009-45. PASJ 1998; 50: 667-673.
% % % 
% % % 
\bibitem[Lewin et~al.(1976)]{1976Lewin}Lewin W. H. G., Hoffman J. A., Doty J., Liller W.
X-Ray Bursts. IAU Circ. 1976; 2994, 2.
% % % 
% % % 
\bibitem[Liu et~al.(2007)]{2007Liu}Liu Q. Z., van Paradijs J., van den Heuvel E. P. J. 
A catalogue of low-mass X-ray binaries in the Galaxy, LMC, and SMC (Fourth edition). A$\&$A 2007; 469: 807-810.
% % % 
% % % 
\bibitem[Lumb et~al.(2012)]{2012lumb}Lumb D. H., Schartel N., Jansen F. A.
\textit{XMM-Newton} (X-Ray Mulit-Mirror Mission) Observatory. ArXiv e-prints 2012; 1202.1651.
% % % 
% % % 
\bibitem[Maeda et~al.(1996)]{1996Maeda}Maeda Y., Koyama K., Sakano M., Takeshima T., Yamauchi S.
A New Eclipsing X-Ray Burster near the Galactic Center: A Quiescent State of the Old Transient A1742-289. PASJ 1996; 48: 417-423.
% % % 
% % % 
\bibitem[Markwardt et~al.(1998)]{1998Markwardt}Markwardt C. B., Marshall F. E., Swank J., Takeshima T.
XTE J1710$-$281 = 1RXS J171012.3$-$280754. IAU Circ. 1998; 6998, 2.
% % % 
% % % 
\bibitem[Masetti et~al.(2007)]{2007Masetti}Masetti N. et al.
IGR J16194$-$2810: a new symbiotic X-ray binary. A$\&$A 2007; 470: 331-337.
% % % 
% % % 
\bibitem[Mason et~al.(2001)]{2001mason}Mason K. O. et al.
The \textit{XMM-Newton} optical/UV monitor telescope. A$\&$A 2001; 365: L36-L44.
% % % 
% % % 
\bibitem[Mathis et~al.(1977)]{Mathis_etal_1977}Mathis, J. S., Rumpl, W. and Nordsieck, K. H.
The size distribution of interstellar grains. ApJ 1977; 217: 425-433.
% % % 
% % % 
\bibitem[Mitsuda et~al.(1990)]{1990Mitsuda}Mitsuda K., Takeshima T., Kii T.
Dust-Grain Scattering of X-Rays Observed during the Lunar Occultation of a Transient X-Ray Source near the Galactic Center.  ApJ 1990; 353, 480.
% % % 
% % % 
\bibitem[Muñoz-Darias et~al.(2009)]{2009_Munozetal}Muñoz-Darias T. et al. Dynamical constraints on the neutron star mass in EXO 0748-676. MNRAS; 394: L136-L140.
% % % 
% % % 
\bibitem[Nespoli et~al.(2010)]{2010Nespoli}Nespoli E., Fabregat J., Mennickent R. E.
K-band spectroscopy of IGR J16358$-$4726 and IGR J16393-4643: two new symbiotic X-ray binaries.  A$\&$A 2010; 516, A94.
% % % 
% % % 
\bibitem[Ozel et~al.(2012)]{Ozel_etal2012}Ozel F, Psaltis D, Narayan R and Santos Villarreal A.
On the Mass Distribution and Birth Masses of Neutron Stars. ApJ 2012; 757, 55.
% % %
% % %
\bibitem[Parmar et~al.(1985)]{1985Parmar}Parmar A. N., Gottwald M., Haberl F., Giommi P., White N. E.
Recent Results on Cataclysmic Variables. The Importance of IUE and Exosat Results on Cataclysmic Variables
and Low-Mass X-Ray Binaries, Burke W. R., ed. EXO:0748$-$676 - an Exciting New X-Ray Transient. 
Recent Results on Cataclysmic Variables. In: Burke W. R., editor. Recent Results on Cataclysmic Variables. 
The Importance of IUE and Exosat Results on Cataclysmic Variables and Low-Mass X-Ray Binaries, Proceedings of an ESA workshop 1985, 236, 119.
% % % 
% % % 
\bibitem[Parmar et~al.(1986)]{1986Parmar_etalPpaper1}Parmar A. N., White N. E., Giommi P.,  Gottwald M.
The Discovery of 3.8 Hour Periodic Intensity Dips and Eclipses from the Transient Low-Mass X-Ray Binary EXO 0748$-$676. ApJ 1986; 308, 199.
% % % 
% % % 
\bibitem[Paul et~al.(2012)]{2012Paul}Paul B., Archana M., Saripalli L.
Simultaneous X-ray and optical observations of thermonuclear bursts in the LMXB EXO 0748$-$676. Bulletin of the Astronomical Society of India 2012; 40: 93-104.
% % % 
% % % 
\bibitem[Ponti et~al.(2019)]{Ponti_etal_2019}Ponti, G. et al.
Evolution of the disc atmosphere in the X-ray binary MXB 1659$-$298, during its 2015-17 outburst. MNRAS 2019; 487: 858-870.
% 
% % % 
% % % 
\bibitem[Ponti et~al.(2018)]{2018Ponti}Ponti G. et al.
\textit{NuSTAR} + \textit{XMM-Newton} monitoring of the neutron star transient AX J1745.6$-$2901.  MNRAS 2018; 473: 2304-2323.
% % % 
% % % 
\bibitem[Ponti et~al.(2017)]{2017Ponti}Ponti G., De K., Muñoz-Darias T., Stella L., Nandra K.
The puzzling orbital period evolution of the LMXB AX J1745.6$-$2901. MNRAS 2017; 464: 840-849.
% % % 
% % % 
\bibitem[Ponti et~al.(2015)]{2015ponti_Fe}Ponti G. et al.
On the Fe K absorption - accretion state connection in the Galactic Centre neutron star X-ray binary AX J1745.6$-$2901. MNRAS 2015; 446: 1536-1550.
% % % 
% % % 
\bibitem[Ponti et~al.(2014)]{2014Ponti}Ponti G., Muñoz-Darias T., Fender R. P.
A connection between accretion state and Fe K absorption in an accreting neutron star: black hole-like soft-state winds? MNRAS 2014; 444: 1829-1834.
% % % 
% % % 
\bibitem[Psaradaki et~al.(2018)]{2018Psaradaki}Psaradaki I., Costantini E., Mehdipour M., Diaz Trigo M.
Modelling the disc atmosphere of the low mass X-ray binary EXO 0748$-$676. A$\&$A 2018; 620. A129.
% % % 
% % % 
\bibitem[Raman et~al.(2018)]{2018raman}Raman G., Maitra C., Paul B.
Observation of variable pre-eclipse dips and disc windsin the eclipsing LMXB XTE J1710$-$281. MNRAS 2018; 477: 5358-5366.
% % % 
\bibitem[R{\'o}{\.z}a{\'n}ska et~al.(2014)]{Roza_etal_2014}Różańska A., Madej J., Bagińska P. ,Hryniewicz K., $\&$ 
Handzlik B. Disk emission and atmospheric absorption lines in black hole candidate 4U 1630$-$472.  A$\&$A  2014; 562, A81. 
% % % 
\bibitem[Raman et~al.(2017)]{2017Gayathri}Raman G., Paul B.
X-ray and optical orbital modulation of EXO 0748$-$676 : A co-variability study using \textit{XMM-Newton}. New A 2017; 54: 122-139.
% % % 
% % % 
\bibitem[Sidoli et~al.(2001)]{2001Sidoli_MXB}Sidoli L., Oosterbroek T., Parmar A. N., Lumb D., Erd C.
An \textit{XMM-Newton} study of the X-ray binary MXB 1659$-$298 and the discovery of narrow X-ray absorption lines. A$\&$A 2001; 379: 540-550.
% % % 
% % % 
\bibitem[Smith et~al.(2016)]{Smith_etal_2016}Smith, R. K., Valencic L. A. and Corrales L.
The Impact of Accurate Extinction Measurements for X-Ray Spectral Models. ApJ 2016; 818, 143.
% % % 
% % % 
\bibitem[Strüder et~al.(2001)]{2001struder}Strüder L. et al.
The European Photon Imaging Camera on \textit{XMM-Newton}: The pn-CCD camera. A$\&$A 2001; 365: L18-L26.
% % % 
% % % 
\bibitem[Suleimanov et~al.(2003)]{2003Suleimanov}Suleimanov, V., Meyer, F., $\&$ Meyer-Hofmeister, E. 
High efficiency of soft X-ray radiation reprocessing in supersoft X-ray sources due to multiple scattering.
A$\&$A 2003; 401, 1009-1015
% % % 
% % % 
\bibitem[Turner et~al.(2001)]{2001turn}Turner M. J. L. et al.
The European Photon Imaging Camera on \textit{XMM-Newton}: The MOS cameras. A$\&$A 2001; 365, L27-L35.
% % % 
% % % 
\bibitem[Wijnands et~al.(2002)]{Wijnands_etal_2002} Wijnands R. et al.
The Burst Behavior of the Eclipsing Low-Mass X-Ray Binary MXB 1659$-$298. ApJ 2002; 566: 1060-1068.
% 
% 
\bibitem[Wilms et~al.(2000)]{Wilms_McCray_00}Wilms, J., Allen, A. and McCray, R.
On the Absorption of X-Rays in the Interstellar Medium. ApJ  2000; 542: 914-924.
% % % 
% % % 
\bibitem[Wolff et~al.(2009)]{Wolff_etal_2009}Wolff M. T. et al. Eclipse timings of the transient Low-Mass X-Ray Binary EXO 0748$-$676. IV. The Rossi X-ray Timing Explorer Eclipses. ApJS 2009; 183: 156-170.
% % % 
% % %
\bibitem[Wolff et~al.(2005)]{Wolff_etal_2005}Wolff M. T. et al. A Strong X-Ray Burst from the Low-Mass X-Ray Binary EXO 0748$-$676. ApJ 2005; 632: 1099-1103.
% % % 
% % % 
\bibitem[Wolff et~al.(2002)]{Wolff_etal_2002}Wolff M. T. et al. Eclipse Timings of the Low-Mass X-Ray Binary EXO 0748–676. III. Orbital Period Jitter Observed with the Unconventional Stellar Aspect Experiment and the Rossi X-Ray Timing Explorer. ApJ 2002; 575: 384-396, 2002.
% % % 
% % %
\bibitem[Younes et~al.(2009)]{2009Younes}Younes G., Boirin L., Sabra B. An \textit{XMM-Newton} view of the dipping low-mass X-ray binary XTE J1710$-$281. A$\&$A 2009; 1099-1103.
% % % 
% % % 
\bibitem[Zycki et~al.(1994)]{1994Zycki}Zycki P. T., Krolik, J. H., Zdziarski, A. A., $\&$ Kallman, T. R. X-Ray Reflection from Photoionized Media in Active Galactic Nuclei. ApJ 1994; 437, 597
% % % 
% % %
% 
% =============
% 
\end{thebibliography}
\end{document}